\documentclass[aps,prr,twocolumn,showpacs,superscriptaddress,amsmath,amssymb,amsfonts,floatfix,citeautoscript,longbibliography]{revtex4-2}

\usepackage[T1]{fontenc}\usepackage[latin1]{inputenc}
\usepackage{dcolumn,graphicx,color,booktabs,microtype,afterpage} 
\usepackage{sidecap}
\usepackage{times}
\usepackage[colorlinks,plainpages=false,linkcolor=blue,urlcolor=blue,citecolor=blue,pdfpagemode=UseNone,pdfstartview=FitBH]{hyperref}
\usepackage[dvipsnames]{xcolor} 

\usepackage{amsmath, amsfonts, amssymb}
\usepackage{bm,bbm}
\usepackage{mathtools}
\usepackage{xfrac}
\usepackage{float}
\usepackage{graphicx}% Include figure files
\usepackage{stackrel}
\usepackage{enumerate}
\usepackage{multirow}
\usepackage{tikz} % colored text box
\usepackage[normalem]{ulem}
\usepackage{wasysym}
\usepackage[caption=false]{subfig} %disable 'caption'
\usepackage{array, booktabs}
	\newcolumntype{P}[1]{>{\centering\arraybackslash}p{#1}} % control the alignment and width
\usepackage{minibox}

\usepackage{stackengine}
\stackMath

\newcommand{\corr}[1]{\langle #1\rangle}

\newcommand{\bds}[1]{\boldsymbol{#1}}

\newcommand{\vekk}[1]{\boldsymbol{#1}}
\newcommand{\f}[2]{\frac{#1}{#2}}

\newcommand{\commentblock}[1]{} % effectively comment out whatever in '{}'

\newcommand{\lode}{\textcolor{blue}}

\newcommand{\changes}{\textcolor{black}}

\DeclareMathOperator*{\Tr}{Tr}

\begin{document}
%-------------------------------------------------------------------------------------------------------------------------
\title{Human-machine collaboration: ordering mechanism\\ of rank$-2$ spin liquid on breathing pyrochlore lattice}
%---------------------------------------------------------------------
\author{Nicolas Sadoune}
\affiliation{Arnold Sommerfeld Center for Theoretical Physics, LMU Munich, Theresienstr. 37, 80333 M\"unchen, Germany}
\affiliation{Munich Center for Quantum Science and Technology (MCQST), 80799 M\"unchen, Germany}
\author{Ke Liu}
\affiliation{Arnold Sommerfeld Center for Theoretical Physics, LMU Munich, Theresienstr. 37, 80333 M\"unchen, Germany}
\affiliation{Munich Center for Quantum Science and Technology (MCQST), 80799 M\"unchen, Germany}
\affiliation{Hefei National Research Center for Physical Sciences at the Microscale and School of Physical Sciences, University of Science and Technology of China, Hefei 230026, China}
\affiliation{Shanghai Research Center for Quantum Science and CAS Center for Excellence in Quantum Information and Quantum Physics, University of Science and Technology of China, Shanghai 201315, China}
\author{Han Yan}
\affiliation{Theory of Quantum Matter Unit, Okinawa Institute of Science and Technology Graduate University, Onna-son, Okinawa 904-0412, Japan}
\affiliation{Institute for Solid State Physics, University of Tokyo, Kashiwa, 277-8581 Chiba, Japan}
\affiliation{Department of Physics and Astronomy, Rice University, Houston, TX 77005, USA}
\affiliation{Smalley-Curl Institute, Rice University, Houston, TX 77005, USA}
\author{Ludovic D.C. Jaubert}
\affiliation{CNRS, Universit\'e de Bordeaux, LOMA, UMR 5798, 33400 Talence, France}
\author{Nic Shannon}
\affiliation{Theory of Quantum Matter Unit, Okinawa Institute of Science and Technology Graduate University, Onna-son, Okinawa 904-0412, Japan}
\author{Lode Pollet}
\affiliation{Arnold Sommerfeld Center for Theoretical Physics, LMU Munich, Theresienstr. 37, 80333 M\"unchen, Germany}
\affiliation{Munich Center for Quantum Science and Technology (MCQST), 80799 M\"unchen, Germany}
\date{\today}
%--------------------------------------------------------------------
\begin{abstract}
Machine learning algorithms thrive on large data sets of good quality. 
Here we show that they can also excel in a typical research setting with little data of limited quality, through 
an interplay of insights coming from machine, and human researchers.
The question we address is the unsolved problem of ordering out of a spin--liquid phase described by an 
emergent rank--2 $U(1)$ gauge theory, as described by [H. Yan {\it et al.}, Phys. Rev. Lett. \textbf{124}, 127203 (2020)].
Published Monte Carlo simulations for this problem are consistent with a strong first--order 
phase transition, %out of the R2-U1 spin liquid, 
but were too noisy for the form of low-temperature order to be identified. 
Using a highly--interpretable machine learning approach based on a support vector machine with 
a tensorial kernel (TKSVM), we re-analyze this Monte Carlo data, gaining new information about the form 
of order that could in turn be interpreted by traditionally-trained physicists. 
\changes{
We find that the low-temperature ordered phase is a form of magnetic order analogous
to a smectic liquid crystal. 
%hybrid nematic order 
%with emergent $\mathbb{Z}_2$ symmetry, 
%
This arises due to a subtle thermal order-by-disorder mechanism, that can be understood 
from the fluctuations of the tensor electric field of the parent rank-2 gauge theory. }
These results were obtained by a back-and-forth process which closely resembles a collaboration between 
human researchers and machines.
We argue that this ``collaborative'' approach may provide a blueprint for solving other problems that have not yielded to human insights alone.
\end{abstract}
\maketitle

%%%%%%%%%%%%%%%%%%%%%%%%%%%%%%%%%%%%%%%%%%%%%%%%%%%%%%%%%%

%-------------------------------------------------------------------------------------------------------------------------
\section{Introduction}
%-------------------------------------------------------------------------------------------------------------------------

%-------------------------------------------------------------------------------------------------------------------------
%  Generative AI grabs the headlines...
%-------------------------------------------------------------------------------------------------------------------------

The past decade has seen great advances in the application of machine learning 
techniques to complex data sets.
Large language models, trained on hundreds of billions of words of text, are now capable of generating 
human-like text, and correctly answering examinations in many different 
subjects \mbox{\cite{Vaswani2023,Kaplan2020,Brown2020}}.
This ``generative AI'' approach has undoubtedly generated great excitement, and such tools have 
found diverse uses as aids to constructing images, legal documents, and even computer code, 
based on existing models. 
However it is less clear what the impact of machine learning will be in fundamental research 
where, by definition, there are no existing models to follow.   
The fact that 21$^{st}$--century science produces vast quantities of data, coming both from experiment 
and simulation, suggests that is also ripe for the application of machine learning.
But as yet, no simple template has emerged for how researchers 
confronted with intractable problems can best use machine--learning 
tools to unlock them.

%-------------------------------------------------------------------------------------------------------------------------
%  Frustrated magents as a natural application of machine learning 
%-------------------------------------------------------------------------------------------------------------------------

Modern condensed matter physics is notable as a source of both novel physical phenomena, 
and large data sets.
A particularly rich vein of new physics, and big data, is found in frustrated magnets.  
Magnetic materials typically undergo a phase transition into an ordered state at low temperatures, 
with the type of order achieved characterized by the symmetries it breaks \cite{Anderson1994}.
In frustrated magnets, competition between different interactions eliminates conventional magnetic order, 
leading to a dizzying array of new phenomena, 
including spin liquids \cite{balents10a,Savary16b}, 
whose excitations include emergent magnetic monopoles \cite{Castelnovo08a,spinicebook}, 
emergent photons \cite{hermele04a,benton12a,spinicebook,pace21a}, 
and Majorana fermions \cite{Kitaev06a,Hermanns18a}, 
as well as exotic, unconventional forms of magnetic order \cite{Shannon2006}.
The effort to understand these phenomena rests heavily on the large datasets produced
by inelastic neutron scattering (INS) experiments, and by Monte Carlo simulation of  
microscopic models.
\changes{And this collision between complex science and big data makes 
frustrated magnetism a natural candidate for the application of machine learning.}

%\lode{These considerations show that the field of frustrated magnetism is a natural 
% candidate for the application of machine learning. }
%%This makes it a natural candidate for the application of machine learning. 

%%%%%%%%%%%%%%%%%%%%%%%%%%%%%%%%%%%%%%%%%%%%

A wide variety of different machine learning techniques have already been applied 
to frustrated magnets, and closely related problems 
\cite{Carrasquilla2017,Wetzel2017a,Wang2017,Hu2017,Wetzel2017b,Beach2018,Liang2018,Wang2018,Liu19a,Greitemann19a,Zhao2019,Greitemann19b,Decelle2019,Canabarro2019,Olsthoorn2020,Liu2021,Doucet2021,Butler2021,Chen2021,Yu2021,samarakoon-arXiv2021,Lozano-Gomez2022,Samarakoon2022}.
In some cases, where it can be trained without prior assumptions, 
machine learning has proved capable of performing a role similar to a human researcher.
This is particularly true of the analysis of Monte Carlo simulation, where interpretable forms of 
machine learning have proved capable of classifying different magnetic phases ``from scratch''. 
A number of different calculations of this type have been carried out for frustrated models 
 \cite{Carrasquilla2017,Liu19a,Greitemann19a,Greitemann19b,Olsthoorn2020,Lozano-Gomez2022}, 
successfully reproducing complicated phase diagrams for systems with emergent gauge 
symmetry \cite{Greitemann19a,Olsthoorn2020}, and even of identifying magnetically ordered 
phases not previously anticipated by other means \cite{Liu2021}.
Meanwhile, on the experimental side, machine learning techniques have been successfully combined with
simulations of dynamics, to extract the parameters of a microscopic models 
from inelastic neutron scattering experiments~\cite{Samarakoon2022}.

%%%%%%%%%%%%%%%%%%%%%%%%%%%%%%%%%%%%%%%%%%%%

One factor which all of these success stories have in common is access to high quality data, 
of the kind which is not always available at the frontiers of understanding.
Precisely because of the wide range of possible outcomes, numerical simulations of frustrated magnets 
are notoriously difficult to perform, showing a strong tendency to fall out of equilibrium at low temperatures \cite{melko04a,jaubert09a,cepas12a,Kato15a,udagawa16a,rau16b,Nekrashevich22a,Fan24a}.
Even where simulations converge, results can still be challenging to interpret \cite{Zhitomirsky08a,Chern13b,Takatsu16a,Taillefumier17a,Hemmatzade24a,ran24a,szabo24a}.
As a consequence, many interesting questions about the emergent phenomena found in frustrated models 
at low temperature remain out of reach.
Can machine learning contribute in this case~?
And if so, how~?

%%%%%%%%%%%%%%%%%%%%%%%%%%%%%%%%%%%%%%%%%%%
%  Fig.1 - lattice and (semi-)ordered ground state 
%%%%%%%%%%%%%%%%%%%%%%%%%%%%%%%%%%%%%%%%%%%

\begin{figure}[t]
\centering
\subfloat[\label{fig:quadpod}]{\includegraphics[width=0.495\columnwidth]{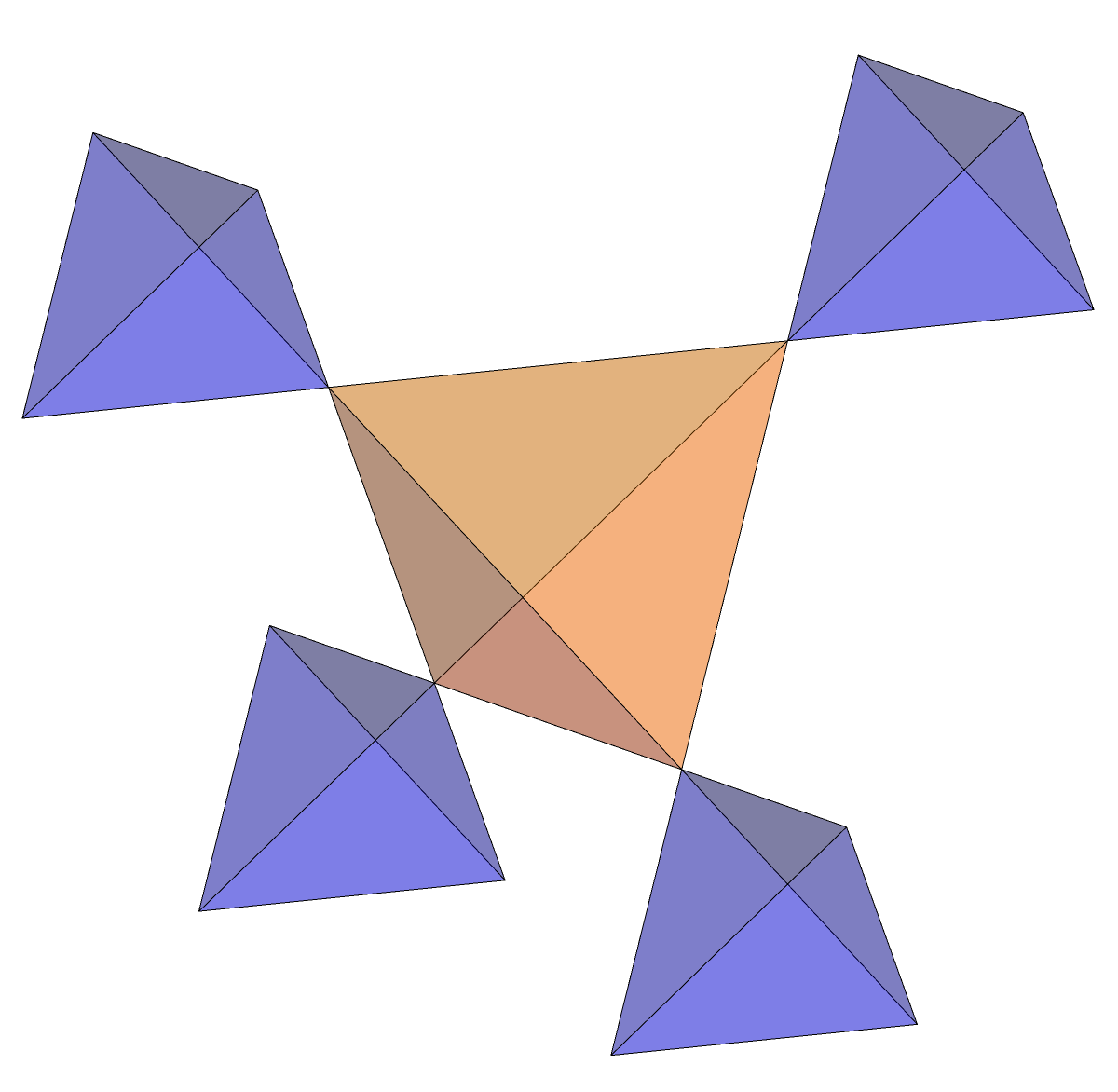}}\quad
\subfloat[\label{fig:GS}]{\includegraphics[width=0.355\columnwidth]{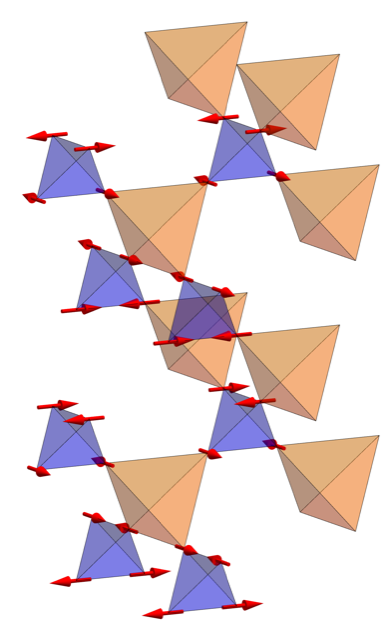}}
\caption{Breathing pyrochlore lattice, and example of ground state found with 
the aid of machine--learning (AI) techniques.
(a) 16--site cubic unit cell of the breathing pyrochlore lattice, with face centered 
cubic (FCC) symmetry.     
%
% The model simulated, Eq.~(\ref{eq:H}), is a Heisenberg antiferromagent with Dzyaloshinskii-Moriya (DM) 
% interactions on $A$--sublattice tetrahedra.  
%
Four small $A$--sublattice tetrahedra, and one large $B$--sublattce tetrahedron, are shown.
(b) Example of magnetic ground state found through the use of AI.
The state shown has a 32--site unit cell, and is associated with wavevector ${\bf q} = W$.
}
\label{fig:lattice}
\end{figure}

%-------------------------------------------------------------------------------------------------------------------------
%  In this paper we...  
%-------------------------------------------------------------------------------------------------------------------------

With these questions in mind, this paper has two objectives: 
firstly, to solve a clearly--motivated physics problem, in an active field of research, 
which has so far proved intractable 
by conventional methods, in part because of the 
failure of simulations to converge.
And, secondly, to show how this solution becomes possible through an interaction between 
human and machine learning.  
% which in many ways resembles a collaboration.

%-------------------------------------------------------------------------------------------------------------------------
%  Emergent higher rank gauge theories as an interesting problem  
%-------------------------------------------------------------------------------------------------------------------------

The problem we consider is an archetypical ``hard problem'' in frustrated magnetism:  
identifying the new phase which arises upon cooling from 
a spin liquid described by an emergent rank$-2$ gauge field, 
in a model motivated by breathing--pyrochlore magnets \cite{Yan20a,Han2022,Zhang2022}. 
The magnetic moments in this spin liquid are disordered, but very 
far from random, satisfying a local constraint which takes the form of a generalized 
Gauss' law  \cite{pretko17a,prem18a,rasmussen2016}.  
This Gauss' law, which is expressed in terms of tensor fields, implies an emergent 
gauge symmetry, with attendant conservation laws.
As a consequence excitations have ``fracton'' character, 
moving in reduced dimension \cite{pretko17a}, 
and characteristic power--laws are found in spin correlations \cite{prem18a,Yan20a}.  

%%%%%%%%%%%%%%%%%%%%%%%%%%%%%%%%%%%%%%%%%%%%

Higher rank--spin liquids are interesting as examples of novel phases of matter, 
but also significant because of deep connections 
with quantum computing \cite{haah2011PhysRevA,Aitchison2024PhysRevB}, 
and attempts to understand quantum gravity  \cite{Cenke2006PhysRevB,Pretko2017PhysRevD,Yan2019PhysRevBfracton1,Yan2019PhysRevBfracton2,Yan2020PhysRevB,yan2023arxivpadic}.
Unfortunately, higher-rank spin liquids are also intrinsically challenging to simulate, 
since they are built from an exponentially large number of degenerate spin 
configurations, which are not connected by the updates used in conventional MC simulation.  
And this challenge gets even greater where the spin liquid undergoes a low--temperature 
transition into a phase in which this degeneracy is wholly or partially lifted, splitting the 
ground state manifold into disjoint pieces.

%%%%%%%%%%%%%%%%%%%%%%%%%%%%%%%%%%%%%%%%%%%
% Fig. 1 - phase diagram
%%%%%%%%%%%%%%%%%%%%%%%%%%%%%%%%%%%%%%%%%%%

\begin{figure}[t]
	\centering
	\includegraphics[width=0.8\columnwidth]{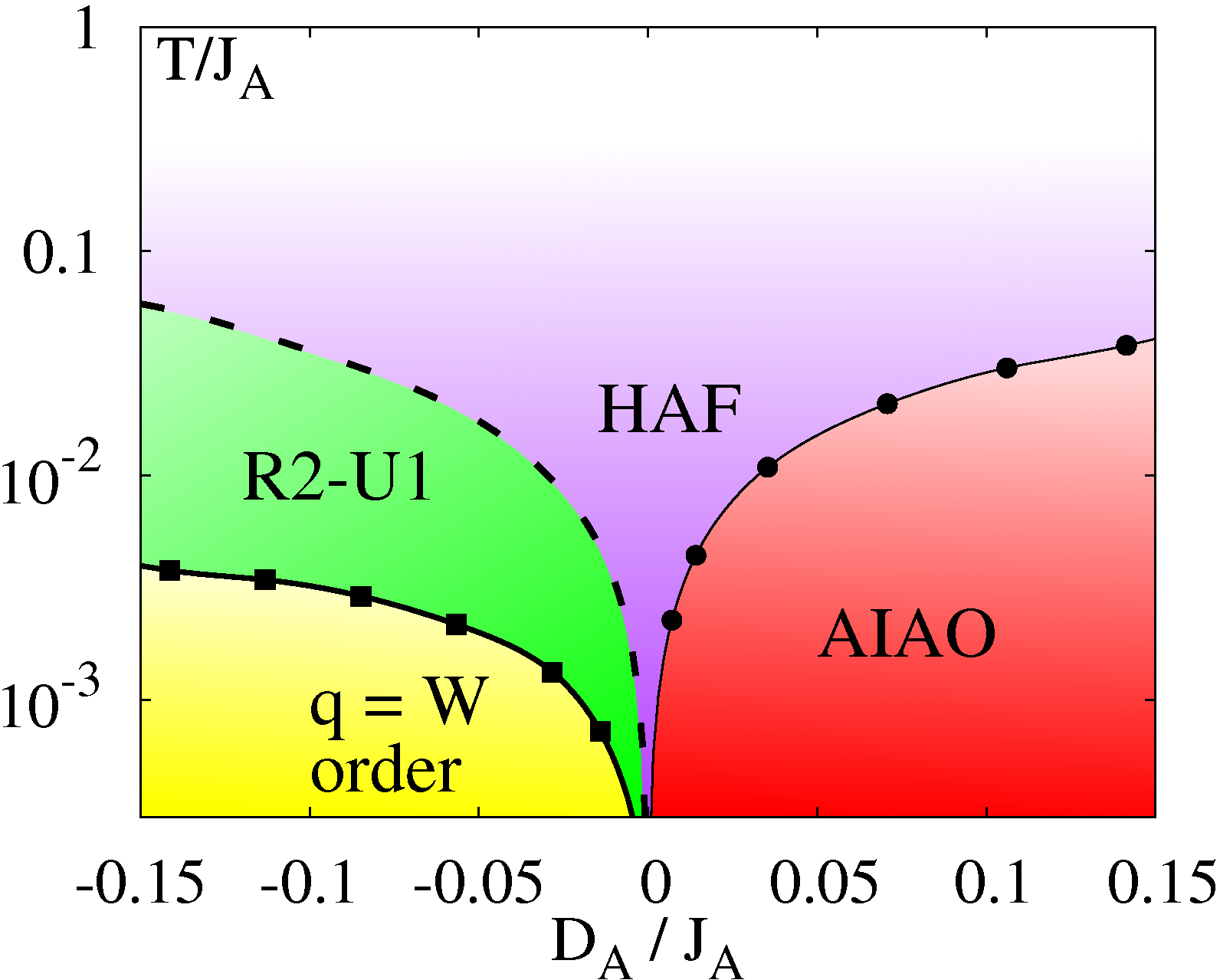}
	\caption{\changes{
	Finite--temperature phase diagram of the breathing pyrochlore model, 
	Eq.~\eqref{eq:H}, as a function of DM interaction $D$,  
	reproduced from \cite{Yan20a}.
	The crossover between the \mbox{R2--U1} spin liquid, %[shaded green], 
	and the rank--1 spin liquid (HAF), 
	% $U(1)\times U(1)\times U(1)$ spin liquid (HAF) %[shaded purple] 
	is shown with a dashed line.
	The thin solid line indicates a continuous transition into all--in all--out 
	order (AIAO).
	The unidentified phase at low temperatures is labelled ${\bf q} = W$ order, 
	and is separated from the   \mbox{R2--U1} spin by a discontinuous 
	phase transition.
	%[shaded yellow].
	%
	Results are taken from MC simulation of Eq.~(\ref{eq:H}), 
	as described in the text.
}}	\label{fig:phase.diagram}
\end{figure}

%%%%%%%%%%%%%%%%%%%%%%%%%%%%%%%%%
%. Fig. 2 - temperature regimes found in simulation
%%%%%%%%%%%%%%%%%%%%%%%%%%%%%%%%%

\begin{figure}[t]
\centering
\includegraphics[width=\columnwidth]{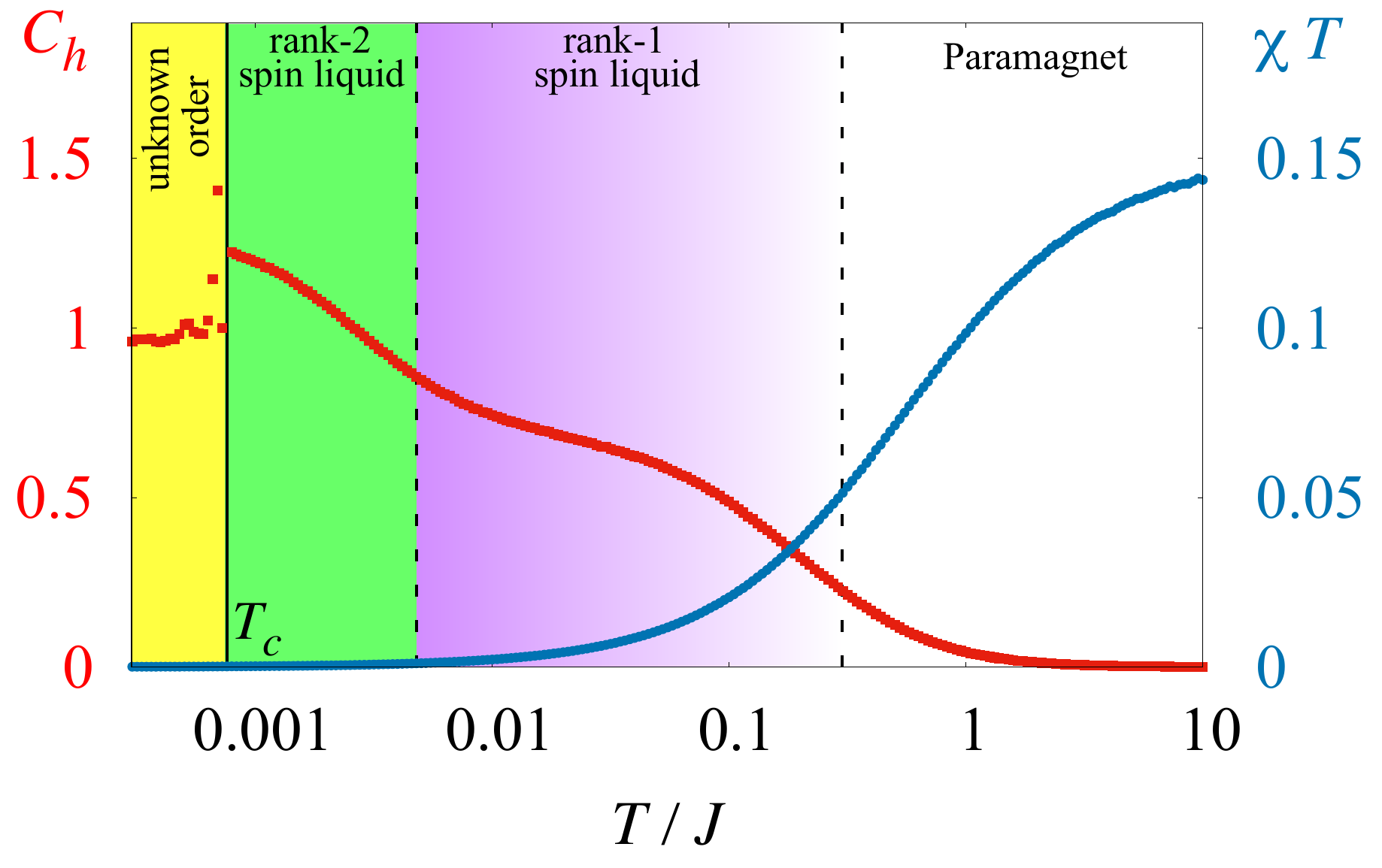}
\caption{
Detail of phase diagram for parameters considered in this Article, 
showing how a rank-2 spin liquid undergoes a phase transition into an 
unidentified form of magnetic order at low temperature.  
Successive crossovers from a high-temperature paramagnet into rank--1 and \mbox{rank--2} 
spin liquids are accompanied by steep rises in the heat capacity $C_h$, with a corresponding 
evolution of the magnetic susceptibility 
$\chi$, % = \frac{1}{NT}(\langle \mathbf{M}^2\rangle-\langle |\mathbf{M}|\rangle^2)$, %[Eq.~(\ref{eq:chi})], 
here plotted as $\chi\times T$.  
 The ordering transition is visible as a sharp discontinuity in $C_h$ for $T \approx 0.001\ J$.
Results are taken from Classical Monte Carlo (MC) simulations of Eq.~(\ref{eq:H}) for a cluster 
of linear dimension $L=8$ ($N=8192$ sites), with parameters Eq.~(\ref{eq:parameters}), 
%with $D=-0.0141\ J$, 
as described in Section~\ref{sec:model} and Section~\ref{sec:MC}. 
%
% The magnetic susceptibility  $\chi$ is defined 
% as $\frac{1}{NT}(\langle \mathbf{M}^2\rangle-\langle |\mathbf{M}|\rangle^2)$ 
% where $\mathbf{M}$ is the total magnetisation.
}
\label{fig:ChChiT}
\end{figure}

%%%%%%%%%%%%%%%%%%%%%%%%%%%%%%%%%%%%%%%%%%%

\begin{figure}[ht]
\centering
\includegraphics[width=0.9\columnwidth]{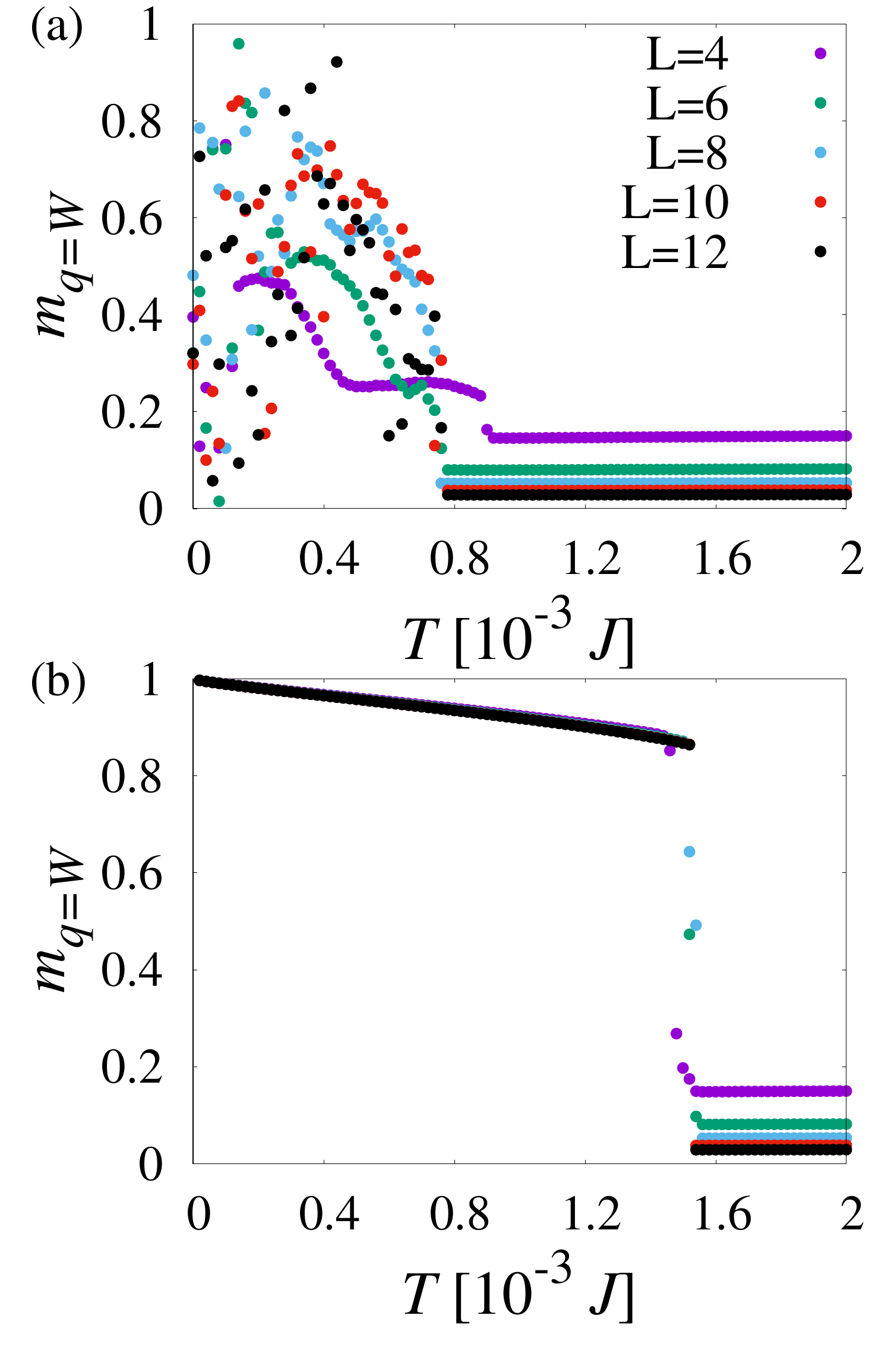}
\caption{
Classical Monte Carlo (MC) simulation results for breathing pyrochlore model before 
and after input from machine learning (AI).
(a)~MC results before input from AI, as discussed by Yan {\it et al.} \cite{Yan20a}.
The staggered magnetization $m_{\bf q}$, at wavevector \mbox{${\bf q} = W$}, 
shows signs of a phase transition into an ordered state for $T \sim 10^{-3}\ \text{J}$.
However simulations at these temperatures %fall out of equilibrium, and 
are too poorly--converged for the form of order to be identified.   
(b)~MC after input from AI, as described in Section~\ref{sec:MC2}.
Once simulations are initialized within the states identified by AI, a clear first--order phase can be 
identified at $T_c \approx 1.5 \times 10^{-3}\ \text{J}$. 
Simulations were carried out for the model Eq.~(\ref{eq:H}), for clusters with linear 
dimension $L \in \{4,6,8,10,12\}$, with parameters Eq.~(\ref{eq:parameters}), as described in Section~\ref{sec:MC}.
Wavevector \mbox{${\bf q} = W$} corresponds to $[1\frac{1}{2}0]$ and symmetry equivalents.
}
\label{fig:MCmqw}
\end{figure}

In earlier work by Yan {\it et al.} [\onlinecite{Yan20a}], a rank--2 U(1) spin liquid was 
identified in classical Monte Carlo (MC) simulation of a model 
\begin{eqnarray}
\label{eq:H}
\mathcal{H}=J\sum_{\langle ij\rangle} \vekk{S}_i\cdot \vekk{S}_j 
+ D \sum_{\langle ij\rangle\in A} \hat{\vekk{d}}_{ij} \cdot (\vekk{S}_i\times\vekk{S}_j).
\end{eqnarray}
in which antiferromagnetic (AF) Heisenberg exchange $J$ competes with Dzyaloshinskii-Moriya (DM) 
interactions $D$, on a breathing pyrochlore lattice [Fig.~\ref{fig:lattice}].
Traditional methods, based on the calculation of structure factors, were used to identify
correlations characteristic of rank--2 spin liquid at temperatures $T_c \lesssim 0.01\ J$ [Fig.~\ref{fig:ChChiT}].
On further cooling, the model exhibited a transition into an ordered state at $T \approx 0.001\ J$.
However, despite the application of a combination of powerful MC techniques 
(heatbath, parallel-tempering and overrelaxation) techniques, it proved impossible 
to thermalize simulations  at these temperatures well enough to identify the form 
of order realized [Fig.~\ref{fig:MCmqw}(a)].
Both the extreme difficulty of simulating this phase transition, 
and the novelty of the physics involved, 
make it an interesting test case for the application of machine learning.  

%%%%%%%%%%%%%%%%%%%%%%%%%%%%%%%%%
%. Fig. 3 - MC results before AI input
%%%%%%%%%%%%%%%%%%%%%%%%%%%%%%%%%

%%%%%%%%%%%%%%%%%%%%%%%%%%%%%%%%%

Here we revisit those simulations, and show what is gained by analyzing data using a variant of a 
support vector machine with tensorial kernel (TKSVM)~\cite{Liu19a,Greitemann19b,Greitemann19a}.    
This highly--interpretable form of machine learning is employed without prior training 
or supervision.
Through a step-by-step iteration of the insights derived from human and machine learning, 
we are able to successfully identify both the form of order, and the mechanism through 
which it occurs.
Ultimately, this permits us to obtain well--equilibrated MC simulation results for the ordered 
phase, by initializing simulations from one of the spin configurations identified by the TKSVM, 
with results shown in Fig.~\ref{fig:MCmqw}.

%%%%%%%%%%%%%%%%%%%%%%%%%%%%%%%%%

The outcomes are twofold:

%%%%%%%%%%%%%%%%%%%%%%%%%%%%%%%%%

Firstly, we are able to characterize the phase found at low temperatures as an ordered phase 
which breaks spin--rotation symmetry through both dipolar and quadrupolar order parameters. 
This phase is entropically selected from a much larger set of degenerate spin configurations, 
and can be identified with fluctuations of the electric field within the rank-2 spin liquid 
\changes{(defined in Section~\ref{sec:fundamentals})}.
We find that the states found at low temperature form a sub--extensive manifold, 
responsible for the quadrupolar order.  
The highest--symmetry state within this manifold, illustrated in Fig.~\ref{fig:GS}, 
has a  32--site unit cell.
This state has a finite staggered magnetization at $\mathbf{q}=W$, 
cf. Fig.~\ref{fig:MCmqw}(b).   

%%%%%%%%%%%%%%%%%%%%%%%%%%%%%%%%%

Secondly, we gain insight into the process through which these results were obtained, 
a back--and--forth dialogue between human and machine insights.    
We argue that this can best be characterized as ``collaboration'', in which neither human 
nor machine plays a dominant role, suggesting one possible paradigm for the solution 
of other difficult problems.
%

%-------------------------------------------------------------------------------------------------------------------------
%  The story so far...
%-------------------------------------------------------------------------------------------------------------------------

The remainder of the article is structured as follows:

%%%%%%%%%%%%%%%%%%%%%%%%%%%%%%%%%

In Section~\ref{sec:fundamentals} we provide details of the model studied and parameters used in simulation, 
\changes{and review the theory of the rank--2 $U(1)$ spin liquid developed in \cite{Yan20a}}.

%%%%%%%%%%%%%%%%%%%%%%%%%%%%%%%%%

In Section~\ref{sec:MC} we summarise the MC results obtained without input from Machine Learning, 
showing how simulations fall out of equilibrium for $T  \lesssim 10^{-3}\ J$, at the phase 
transition from a rank--2 spin liquid into an unidentified form of magnetic order.

%%%%%%%%%%%%%%%%%%%%%%%%%%%%%%%%%

In Section~\ref{sec:ML} we describe application of a Support Vector Machine with 
Tensorial Kernel (TKSVM) to MC data, and sumarise results found for 
kernels of rank-1 and rank-2.   
Using this information, we reconstruct the spin configurations found on A-- and B--sublattice 
tetrahedra for $T < T_c$. 

%%%%%%%%%%%%%%%%%%%%%%%%%%%%%%%%%

In Section~\ref{sec:MC2} we revisit MC simulations in the light of what we have learned from
the TKSVM, quenching from an ordered state with 32--site unit cell, built from 
A-- and B--sublattice tetrahedra with configurations identified in Section~\ref{sec:ML}.
Simulations then converge, providing a new estimate $T_c  \approx 1.5\ \times 10^{-3}\ J$.

%%%%%%%%%%%%%%%%%%%%%%%%%%%%%%%%%

In Section~\ref{sec:Z2symmetry} we explore the further implications of the correlations 
identified in  Section~\ref{sec:ML}, identifying an emergent $\mathbb{Z}_2$ symmetry, associated 
with flipping an entire plane of spins, and the 
associated manifold of $2^L$ degenerate ground states.

%%%%%%%%%%%%%%%%%%%%%%%%%%%%%%%%%

In Section~\ref{sec:Td} we reexpress what we have learned about spin configurations for $T < T_c$
in the language of irreps of $T_d$, providing a new interpretation of the  $\mathbb{Z}_2$ symmetry identified 
in Section~\ref{sec:Z2symmetry}.   

%%%%%%%%%%%%%%%%%%%%%%%%%%%%%%%%%

We conclude in Section~\ref{sec:conclusion} with a summary of results, and reflect 
on the interplay of human and machine insights which made it possible to identify 
the form of ground state order in this model.   
We identify a number of other problems which might be susceptible to a similar approach.

%%%%%%%%%%%%%%%%%%%%%%%%%%%%%%%%%

Further technical details are provided in three Appendices.

%%%%%%%%%%%%%%%%%%%%%%%%%%%%%%%%%

In Appendix~\ref{sec:Def} we summarise the conventions used in defining 
the breathing pyrochlore lattice and irreps of $T_d$.

%%%%%%%%%%%%%%%%%%%%%%%%%%%%%%%%%

In Appendix~\ref{sec:TKSVM.overview} we provide an concise 
overview of the TKSVM approach used to analyze MC data.

%%%%%%%%%%%%%%%%%%%%%%%%%%%%%%%%%

Finally, in Appendix~\ref{sec:PatternInterpretation} we discuss the way in which 
decision functions obtained in TKSVM can be interpreted in terms of spin correlations.

%%%%%%%%%%%%%%%%%%%%%%%%%%%%%%%%%%%%%%%%%%%%%%%%%%%%%%%%%%
\changes{\section{Breathing pyrochlore model and known phases}}
%%%%%%%%%%%%%%%%%%%%%%%%%%%%%%%%%%%%%%%%%%%%%%%%%%%%%%%%%%
\label{sec:fundamentals}

%%%%%%%%%%%%%%%%%%%%%%%%%%%%%%%%%%%%%%%%%%%%%%%%%%%%%%%%%%
\subsection{Model}
%%%%%%%%%%%%%%%%%%%%%%%%%%%%%%%%%%%%%%%%%%%%%%%%%%%%%%%%%%
\label{sec:model}

The model we consider is a classical Heisenberg anti-ferromagnet on the breathing pyrochlore lattice, 
with competing Dzyaloshinskii-Moriya (DM) interactions on A--sublattice tetrahedra, 
\begin{eqnarray}
\mathcal{H}=J\sum_{\langle ij\rangle} \vekk{S}_i\cdot \vekk{S}_j 
+ D \sum_{\langle ij\rangle\in A} \hat{\vekk{d}}_{ij} \cdot (\vekk{S}_i\times\vekk{S}_j) \;,
\nonumber
\end{eqnarray}
previously defined in Eq.~(\ref{eq:H}), and illustrated in Fig.~\ref{fig:lattice}.  
The position of lattice sites, and orientation of the DM vectors $\hat{\vekk{d}}_{ij}$, 
are specified in Appendix~\ref{sec:Def}.

%%%%%%%%%%%%%%%%%%%%%%%%%%%%%%%%%%%%%%%%%%%%

This model was studied by Yan {\it et al.}~\cite{Yan20a}, as a route to realizing an 
exotic rank$-2$ $U(1)$ spin liquid, reviewed below.

The parameters $J > 0, D<0$ are also applicable to A--sublattice tetrahedra 
within the breathing pyrochlore magnet  
${\rm Ba}_3{\rm Yb}_2{\rm Zn}_5\rm{O}_{11}$~\cite{Haku16a,Rau16a,Chern22a}.
However in this case the interactions on the $B$--sublattice tetrahedra 
are two small to accurately characterize. 

%%%%%%%%%%%%%%%%%%%%%%%%%%%%%%%%%%%%%%%%%%%%

In what follows we breifly recap what is already known about the different orderd 
and spin liquid phases supported by this model, as summarised in Fig.~\ref{fig:phase.diagram}.

%%%%%%%%%%%%%%%%%%%%%%%%%%%%%%%%%%%%%
\subsection{Ordered ground state for $D > 0$}
%%%%%%%%%%%%%%%%%%%%%%%%%%%%%%%%%%%%%

\changes{
For positive $D>0$, the ground state is a simple all-in all-out ordered 
phase~\cite{canals08a,chern10b,noculak23a}.  
This state is achieved through a first order phase transition, 
denoted with a solid line in Fig.~\ref{fig:phase.diagram}.
}

%%%%%%%%%%%%%%%%%%%%%%%%%%%%%%%%%%%%%%%%%%%%%%%%%%%%%%%%%%
\subsection{Rank--1 spin liquid for $D=0$ [standard formulation]}
%%%%%%%%%%%%%%%%%%%%%%%%%%%%%%%%%%%%%%%%%%%%%%%%%%%%%%%%%%
\label{sec:U1.U1.U1}
\changes{
For \mbox{$J_B = J_A =J=1$} and \mbox{$D=0$} the model reduces to a Heisenberg 
anti-ferromagnet (HAF) on the pyrochlore lattice, which provides a celebrated example of classical 
spin liquid~\cite{Moessner98c,Moessner1998-PRB58}.
In the large--N limit \cite{isakov04b}, this spin liquid can be understood as three, independent, copies 
of a spin--ice--like $U(1)$ spin liquid.
We briefly review this argument below.   
A more extended discussion can be found in \cite{henley05a}.
}

%%%%%%%%%%%%%%%%%%%%%%%%%%%%%%%%%%%%%%%%%%%%
% Table 1 - irreps of T_d
%%%%%%%%%%%%%%%%%%%%%%%%%%%%%%%%%%%%%%%%%%%%

\begin{table*}[!htb]
\hspace*{-1cm}
\centering
    \begin{tabular}{ c  c }
        \hline
        \hline
        irrep & definition in terms of spin components \\
        \hline
         \vspace*{0.2cm}
%         $m_{A_1}$               & $\frac{1}{2\sqrt{3}}(
%                                  S_0^x + S_0^y + S_0^z + 
%                                   S_1^x + S_1^y + S_1^z + 
%                                   S_2^x + S_2^y + S_2^z + 
%                                   S_3^x + S_3^y + S_3^z)$ \\
        $m_\mathsf{A_2}$               & $\frac{1}{2\sqrt{3}}(
                                  S_0^x + S_0^y + S_0^z + 
                                   S_1^x - S_1^y - S_1^z - 
                                   S_2^x + S_2^y - S_2^z - 
                                   S_3^x - S_3^y + S_3^z)$ \\
        \vspace*{0.2cm}
        $\mathbf{m}_\mathsf{E}$          & $\begin{pmatrix}
                                        \frac{1}{2\sqrt{6}}(
                                            -2S_0^x + S_0^y + S_0^z -
                                             2S_1^x - S_1^y - S_1^z + 
                                             2S_2^x + S_2^y - S_2^z + 
                                             2S_3^x - S_3^y + S_3^z) \\
                                         \frac{1}{2\sqrt{2}}(
                                            - S_0^y + S_0^z 
                                             + S_1^y - S_1^z 
                                             - S_2^y - S_2^z 
                                             + S_3^y + S_3^z)
                                   \end{pmatrix}$ \\
        \vspace*{0.2cm}
        $\mathbf{m}_\mathsf{T_{1+}}$   & $\begin{pmatrix}
                                        \frac{1}{2}(
                                            S_0^x + 
                                             S_1^x + 
                                             S_2^x + 
                                             S_3^x)  \\
                                        \frac{1}{2}(
                                            S_0^y + 
                                             S_1^y + 
                                             S_2^y + 
                                             S_3^y)  \\
                                        \frac{1}{2}(
                                            S_0^z + 
                                             S_1^z + 
                                             S_2^z + 
                                             S_3^z)
                                   \end{pmatrix}$  \\
        \vspace*{0.2cm}
        $\mathbf{m}_\mathsf{T_{1-}}$   & $\begin{pmatrix}
                                    \frac{-1}{2\sqrt{2}}(
                                            S_0^y + S_0^z 
                                           - S_1^y - S_1^z  
                                           - S_2^y + S_2^z  
                                           + S_3^y - S_3^z)  \\
                                    \frac{-1}{2\sqrt{2}}(
                                            S_0^x + S_0^z 
                                           - S_1^x + S_1^z  
                                           - S_2^x - S_2^z  
                                           + S_3^x - S_3^z)  \\
                                    \frac{-1}{2\sqrt{2}}(
                                            S_0^x + S_0^y 
                                           - S_1^x + S_1^y  
                                           + S_2^x - S_2^y  
                                           - S_3^x - S_3^y)  \\
                                   \end{pmatrix}$ \\
        \vspace*{0.2cm}
        $\mathbf{m}_\mathsf{T_2}$      & $\begin{pmatrix}
                                    \frac{1}{2\sqrt{2}}(
                                          - S_0^y + S_0^z 
                                           + S_1^y - S_1^z  
                                           + S_2^y + S_2^z  
                                           - S_3^y - S_3^z)  \\
                                    \frac{1}{2\sqrt{2}}(
                                            S_0^x - S_0^z 
                                           - S_1^x - S_1^z  
                                           - S_2^x + S_2^z  
                                           + S_3^x + S_3^z)  \\
                                    \frac{1}{2\sqrt{2}}(
                                          - S_0^x + S_0^y 
                                           + S_1^x + S_1^y  
                                           - S_2^x - S_2^y  
                                           + S_3^x - S_3^y)  \\
                                   \end{pmatrix}$  \\
        \hline
  \end{tabular}
\caption{Irreps of the symmetry group of isolated tetrahedron, $T_d$. 
The convention for labelling sites is defined in Appendix~\ref{sec:Def}
        }
\label{table:irreps.Td}
\end{table*}
 
%%%%%%%%%%%%%%%%%%%%%%%%%%%%%%%%%%%

\changes{
We first consider a single component of spin, 
\begin{eqnarray}
	S^\alpha_i \; , \quad 
	\alpha \in  \{x, y, z \}  \; .  
\end{eqnarray}
Within any given tetrahedron, this contributes to the Hamiltonian as 
\begin{eqnarray}
	\Delta {\mathcal H}^\alpha_{\sf tet} = J \left(\sum_{i \in \sf{tet}} S_i^\alpha\right)^2 \; .
\label{eq:DeltaH}
\end{eqnarray}
The energy associated with Eq.~(\ref{eq:DeltaH}) is minimized by 
spin configurations satisfying an emergent Gauss' law 
\begin{eqnarray}
	\nabla\cdot\bm{E}_i^\alpha = 0 \; , \qquad  \bm{E}_i^{\alpha} = S_i^\alpha \bm{\hat{z}}_i \; , 
	\label{eqn:Gauss.law.U1}
\end{eqnarray}
where $\{ \bm{\hat{z}}_i \}$ \changes{is the local unit vector on site $i$, pointing from one center of a tetrahedron to another.}
%form an orthogonal set of unit vectors. 
%
The emergent electric field $\bm{E}_i^{\alpha}$ is defined on each site $i$ 
of the pyrochlore lattice, or equivalently, each link of the dual diamond lattice \cite{henley05a}.
}
\changes{
	Note that here, although $\alpha$ labels the component of spin $\bm{S}$, but on $\bm{E}$ it labels the ``copy'' of electric field. There are three copies of electric field, each obeying its own Gauss's law.} 

%%%%%%%%%%%%%%%%%%%%%%%%%%%%%%%%%%%%%%%

\changes{
Within a ``large--N'' limit, where the number of spin components $N \to \infty$, 
we can soften the spin--length constraint, 
\begin{eqnarray}
|{\bf S}_i| = 1\ \forall\ i \in \{ 1 \ldots N_{\sf sites} \} 
	\rightarrow %_{N \to \infty} 
	\frac{1}{N_{\sf sites}} \left| \sum_{i=1}^{N_{\sf sites}} {\bf S}_i \right|  = 1 \; .
\end{eqnarray}
% 
% where $N_{\sf sites}$ is the number of sites in the lattice.
%
Under this assumption, the three different components of spin $ \{S^x, S^y, S^z \}$ 
fluctuate independently, and Eq.~(\ref{eqn:Gauss.law.U1}) defines three independent 
copies of a Gauss' law.
It follows that we can introduce three, independent, fields $ {\bf A}^\alpha$ 
with vector (rank--1) character, such that 
\begin{eqnarray}
	\bm{E}_i^{\alpha} = \nabla \times {\bf A}^\alpha \; , 
\end{eqnarray}
and physical properties of the system invariant under the three, independent, 
$U(1)$ gauge transformations  
\begin{eqnarray}
	 {\bf A}^\alpha   \to  {\bf A}^\alpha + \nabla \theta^\alpha  \; . 
\end{eqnarray}
For this reason, the spin liquid found in the Heisenberg AF on the pyrochlore 
lattice is commonly known as a 
\begin{eqnarray}
	U(1) \times U(1) \times U(1) \nonumber  
\end{eqnarray}
spin liquid, with each $\theta^\alpha$ defining an independent $U(1)$ gauge 
degree of freedom~\cite{Henley10a}.   
In what follows, we shall refer to the spin liquid found for $D=0$ as the 
a \textit{rank-1  $U(1)$} spin liquid, to distinguish it 
from the \textit{rank--2} $U(1)$ spin liquid, to be  introduced below.
}

%%%%%%%%%%%%%%%%%%%%%%%%%%%%%%

\changes{
The defining property of the rank--1 $U(1)$ spin liquid is its Gauss's law, 
Eq.~(\ref{eqn:Gauss.law.U1}), which leads to algebraic correlation of spins, 
and is experimentally observable through two--fold  ``pinch--points'' 
in the equal time structure factor $S({\bf q})$ \cite{henley05a}.
This spin liquid persists down to $T\rightarrow 0$ at $D=0$, 
connecting smoothly with the ground state manifold defined by 
\begin{eqnarray}
	\Delta {\mathcal H}^\alpha_{\sf tet} = 0 \quad \forall \quad \alpha \in \{x,y,z \} \; .
\end{eqnarray}
For finite values of $|D|\ll J$, the rank--1 spin liquid survives at finite temperature, 
thanks to the extensive entropy, before crossing over to a conventional paramagnet 
for $T \gg J$ [Fig.~\ref{fig:phase.diagram}].
}

%%%%%%%%%%%%%%%%%%%%%%%%%%%%%%%%%%%%%
\subsection{Rank--1 spin liquid for $D = 0$ [tensor formulation]}
%%%%%%%%%%%%%%%%%%%%%%%%%%%%%%%%%%%%%
\label{sec:meet.the.irreps}

\changes{
It is possible to formulate the rank--1 $U(1)$ spin liquid described above
as a special case of a more general rank--2 tensor theory.
In what follows we review this approach, as preliminary to describing
the rank--2 spin liquid found for \mbox{$D<0$}.
Further details can be found in~\cite{benton16a,Yan20a,BentonThesis,Yan2024PhysRevB,Yan2024PhysRevB2}.
}

%%%%%%%%%%%%%%%%%%%%%%%%%%%%%%%%%%%%%

\changes{
We start with the observation that it is possible to combine the three vector 
fields $\bm{E}^\alpha$ into a single tensor field
\begin{eqnarray}
	E^{\sf HAF}_{\alpha \beta} = 
	\begin{bmatrix}
		 ({E}^x)_x, ({E}^x)_y,  ({E}^x)_z\\
		 ({E}^y)_x, ({E}^y)_y,  ({E}^y)_z\\
		 ({E}^z)_x, ({E}^z)_y,  ({E}^z)_z\\
	\end{bmatrix} \; .
	\label{eq:E.HAF}
\end{eqnarray}
Doing so, the three Gauss' laws, Eq.~(\ref{eqn:Gauss.law.U1}), 
can be collectd in a single equation 
\begin{equation}
	\partial_\beta  E^{\sf HAF}_{\alpha \beta} % \equiv  \partial_\beta (\bm{E}^\alpha)_\beta  
	=  0 \; .
\label{eq:Gauss.law.HAF}
\end{equation}
We can link this tensor formulation of emergent electrostatics to the energetics of the problem 
by transcribing the Hamiltonian, Eq.~(\ref{eq:H}), in terms of the irreps 
\begin{eqnarray}
	\{ \mathbf{m}_\lambda \}_{T_d} 
	&=&   
	\{  {m}_\mathsf{A_2}, \mathbf{m}_\mathsf{E},  \mathbf{m}_\mathsf{T_1+}, \mathbf{m}_\mathsf{T_1-},\mathbf{m}_\mathsf{T_2}  \}
\end{eqnarray}
of the tetrahedral symmetry group $T_d$ [Table~\ref{table:irreps.Td}], to give
\begin{equation}
	\mathcal{H}  = \frac{1}{2} \sum_{A, \lambda} a_{A,\lambda} \mathbf{m}_\lambda^2 + \frac{1}{2} \sum_{B, \lambda} a_{B,\lambda} \mathbf{m}_\lambda^2 \; , 
\label{eq:H.irreps}
\end{equation}
where the sum on $A(B)$ sum runs over all $A(B)$--sublattice tetrahedra, 
and the coefficients $a_{A(B),\lambda}$ are defined in Table~\ref{table:coefficients}.
}

%%%%%%%%%%%%%%%%%%%%%%%%%%%%%%%%%%%%%%%%

\changes{
In case of the Heisenberg antiferromagnet, \mbox{$D=0$}, we have
\begin{equation}
	\label{eqn.u1acondition}
0 < a_\mathsf{A_2} = a_\mathsf{E} = a_\mathsf{T_2} = a_\mathsf{T_{1-}}  < a_\mathsf{T_{1+}}.
\end{equation}
and it follows that rank--1 spin liquid contains fluctuations transforming with the irreps 
\begin{eqnarray}
	\{m_\lambda \}_{\sf HAF} = \{ {m}_\mathsf{A_2}, \mathbf{m}_\mathsf{E}, \mathbf{m}_\mathsf{T_1-},\mathbf{m}_\mathsf{T_2}  \}
\label{eq:irreps.in.HAF}
\end{eqnarray}
while the irrep $\bm{m}_\mathsf{T_{1+}}$ describes excitations with finite charge.
}

%%%%%%%%%%%%%%%%%%%%%%%%%%%%%%%%%%%%%%%%%%%%%%
% Table 2 - coefficients a_\lambda
%%%%%%%%%%%%%%%%%%%%%%%%%%%%%%%%%%%%%%%%%%%%%%

\begin{table}[!t]
	\begin{tabular}{ | c | c | }
		\hline
		coefficient &
		parameterization  \\
		\hline
		$a_{\mathsf{A_2},\text{A}}$   & $  -J- 4D/\sqrt{2}$\\
		$a_{\mathsf{T_2},\text{A}}$   &  $-J -2D/\sqrt{2}$ \\
		$a_{\mathsf{T_{1+}},\text{A}} $  &  $3J$\\
		$a_{\mathsf{T_{1-}},\text{A}} $ &    $-J + 2 D/\sqrt{2}$  \\
		$a_{\mathsf{E},\text{A}} $ &   $-J + 2 D/\sqrt{2}$  \\
		\hline
		$a_{\mathsf{A_2},\text{B} }$ & $ -J$  \\
		$a_{\mathsf{E},\text{B}}$  & $-J$\\
		$a_{\mathsf{T_2},\text{B}}$  &  $-J $\\
		$a_{\mathsf{T_{1-}},\text{B} }$ &    $-J $\\
		$a_{\mathsf{T_{1+}},\text{B} }$  &  $3J$ \\
		\hline
	\end{tabular}
	\caption{Coefficients $a_\lambda$ appearing in Eq.\eqref{eq:H.irreps}, 
		expressed in terms of the parameters of the original 
		Hamiltonian, Eq.\eqref{eq:H}.
	}
	\label{table:coefficients}
\end{table}
%%%%%%%%%%%%%%%%%%%%%%%%%%%%%%%%%%%%%%%%

\changes{
Since the spin configurations within the rank--1 spin liquid are charactertised
by the irreps $\{m_\lambda \}_{\sf HAF}$, it must be possible to 
decompose the emergent electric field $\bm{E}^{\sf HAF}$ [Eq.~(\ref{eq:E.HAF.irreps})] 
in terms of the same set of irreps.
Following \cite{benton16a,Yan20a,BentonThesis}, we write
\begin{equation}
\bm{E}^{\sf HAF} 
	=  \bm{E}^{\sf HAF}_{\sf sym.} 
	+  \bm{E}^{\sf HAF}_ {\sf antisym.}
	+ \bm{E}^{\sf HAF}_{\sf trace}
\label{eq:E.HAF.irreps}
\end{equation}
where we have decomposed the tensor $\bm{E}^{\sf HAF}$ into its symmetric component
\begin{equation}
\bm{E}^{\sf HAF}_{\sf sym.}  = 
\begin{bmatrix}
\frac{2}{\sqrt{3}}m_\mathsf{E}^1  &  m_\mathsf{T_{1-}}^z &  m_\mathsf{T_{1-}}^y  \\
m_\mathsf{T_{1-}}^z  & -\frac{1}{\sqrt{3}}m_\mathsf{E}^1 - m_\mathsf{E}^2  &   m_\mathsf{T_{1-}}^x  \\
m_\mathsf{T_{1-}}^y &  m_\mathsf{T_{1-}}^x   &  -\frac{1}{\sqrt{3}}m_\mathsf{E}^1 + m_\mathsf{E}^2
\end{bmatrix}   \; , 
\label{eq:Ematrix}
\end{equation}
antisymmetric component 
\begin{equation}
( E^{\sf HAF}_ {\sf antisym.})_{ij} = - \epsilon_{ijk} m_\mathsf{T_2}^k \; , 
\end{equation}
and trace
\begin{equation}
	 ({E}^{\sf HAF}_{\sf trace})_{ij}  = -\delta_{ij}\sqrt{\frac{2}{3}}m_\mathsf{A_2} \; .
\end{equation}
}

%%%%%%%%%%%%%%%%%%%%%%%%%%%%%%%%%%%

\changes{
Spatial derivative of the irreps contributing to $\bm{E}^{\sf HAF}$
are linked through the requirement of the continuity of the fields $m_\lambda$ 
 \cite{BentonThesis}. 
 \changes{In particular, we have only $\bm{m}_{\mathsf{T_{1+}}} = 0$ enforced by the Hamiltonian with coefficients of Eq.~\eqref{eqn.u1acondition}, which imposes the conditions }
\begin{eqnarray}
 && 
 \frac{2}{\sqrt{3}}
\begin{bmatrix}
\partial_x  m_\mathsf{E}^1  \\
-\frac{1}{2} \partial_y m_\mathsf{E}^1 - \frac{\sqrt{3}}{2} \partial_y m_\mathsf{E}^2  \\
-\frac{1}{2} \partial_y m_\mathsf{E}^1 + \frac{\sqrt{3}}{2} \partial_y m_\mathsf{E}^2
\end{bmatrix}    
- 
\begin{bmatrix}
\partial_y m_\mathsf{T_{1-}}^z + \partial_z m_\mathsf{T_{1-}}^y  \\
\partial_z m_\mathsf{T_{1-}}^x + \partial_x m_\mathsf{T_{1-}}^z  \\
\partial_x m_\mathsf{T_{1-}}^y + \partial_y m_\mathsf{T_{1-}}^x   
\end{bmatrix} 
\nonumber\\
%%%%%%%%%%%%%%%%%%%%%%5
 && -\sqrt{\frac{2}{3}}\bm{\nabla}m_\mathsf{A_{2}}
+ \bm{\nabla} \times \mathbf{m}_\mathsf{T_{2}}
= 0 \; .
\label{eq:HAF.constraint}
\end{eqnarray}
Once these conditions are taken into account, the field $\bm{E}^{\sf HAF}$ [Eq.~(\ref{eq:E.HAF.irreps})]
satisfies the emergent Gauss' law, Eq.~(\ref{eq:Gauss.law.HAF}).
}

\changes{We finally arrive at a phenomenological Hamiltonian describing the low-energy physics of rank-1 U(1) theory (R1U1) as  
\begin{equation}
\mathcal{H}_{\sf  R1U2}  \sim U \sum_{\alpha=1,2,3} (\partial_\beta  {E}_{\alpha \beta})^2  .
\end{equation}
where $ \bm{E}$ is $\bm{E}^{\sf HAF}$. 
We see here that the Guass's law is enforced energetically, arising from the $a_{\mathsf{T_{1+}}}{\bm m}_{\mathsf{T_{1+}}}^2$ term. 
}
\\

%%%%%%%%%%%%%%%%%%%%%%%%%%%%%%%%%%%%%
\subsection{Rank--2 spin liquid for $D < 0$ [microscopic]}
%%%%%%%%%%%%%%%%%%%%%%%%%%%%%%%%%%%%%

\changes{
For negative \mbox{$D < 0$}, the Hamiltonian Eq.~(\ref{eq:H})  does not select a ordered 
ground state, but instead a (sub)extensive manifold of ground states, 
which are in turn a subset of the manifold of states associated with 
the rank--1 $U(1)$ spin liquid.
As a consequence, rather than undergoing a phase transition into an ordered state 
on cooling, as happens for \mbox{$D > 0$} [Fig.~\ref{fig:phase.diagram}], 
the rank--1 $U(1)$ spin liquid first undergoes a crossover into a new 
form of spin liquid.
The structure of this ground state manifold, and the resulting intermediate 
spin liquid, can be deconstructed using the language of irreps, introduced above.
}

%%%%%%%%%%%%%%%%%%%%%%%%%%%%%%%%%%%%%

\changes{
Considering first the A--sublattice tetrahedra, for  \mbox{$J > 0$}, \mbox{$D < 0$}, 
the coefficients $a_{\text{A}, \lambda}$  satisfy the condition
\begin{equation}
a_{\text{A},\mathsf{E}}   =   a_{\text{A},\mathsf{T_{1-}}}  < 
a_{\text{A},\mathsf{A_2}}  ,\  a_{\text{A}, \mathsf{T_2} }   , \  a_{\text{A},\mathsf{T_{1+}}} \; ,
\label{eqn.a.in.A}
\end{equation}
which implies that only the fields 
\begin{eqnarray}
\{ m_\lambda \}_{\sf D < 0}  = \{ \mathbf{m}_{\mathsf{E}} , \mathbf{m}_{\mathsf{T_{1-}}} \}
\end{eqnarray}
enter into the ground state of Eq.~(\ref{eq:H.irreps}).
Turning to the B--sublattice tetrahedra, we find a simplified form of the equation 
of continuity Eq.~(\ref{eq:HAF.constraint}), via,
\begin{eqnarray}
&& 
 \frac{2}{\sqrt{3}}
\begin{bmatrix}
\partial_x  m_\mathsf{E}^1  \\
-\frac{1}{2} \partial_y m_\mathsf{E}^1 - \frac{\sqrt{3}}{2} \partial_y m_\mathsf{E}^2  \\
-\frac{1}{2} \partial_y m_\mathsf{E}^1 + \frac{\sqrt{3}}{2} \partial_y m_\mathsf{E}^2
\end{bmatrix}    
- 
\begin{bmatrix}
\partial_y m_\mathsf{T_{1-}}^z + \partial_z m_\mathsf{T_{1-}}^y  \\
\partial_z m_\mathsf{T_{1-}}^x + \partial_x m_\mathsf{T_{1-}}^z  \\
\partial_x m_\mathsf{T_{1-}}^y + \partial_y m_\mathsf{T_{1-}}^x   
\end{bmatrix} \nonumber\\
&&= 0 \; .
\label{eq:D.less.0.constraint}
\end{eqnarray}
Eliminating fluctuations of $m_{\sf T_2}$ and $m_{\sf A_2}$ reduces the tensor electric field $\bm{E}$ 
to its symmetric, traceless component $\bm{E}^{\sf HAF}_{\sf sym.}$ [Eq.~(\ref{eq:Ematrix})].
This symmetric, traceless, electric field satisfies a Gauss' law
\begin{equation}
	\partial_\alpha E_{\alpha\beta} = 0
\label{eq:Gausslaw}
\end{equation}
which defines a rank-2 U(1) gauge theory~\cite{pretko20a}.
}

%%%%%%%%%%%%%%%%%%%%%%%%%%%%%%%%%%%%%
\subsection{Rank--2 spin liquid for $D < 0$ [phenomenological]}
%%%%%%%%%%%%%%%%%%%%%%%%%%%%%%%%%%%%%
 
\changes{
The spin liquid defined through Eq.~(\ref{eq:Gausslaw}) has 
properties which are substantially different from the rank--1 $U(1)$
spin liquid discussed in Section~\ref{sec:U1.U1.U1}.
We can gain some insight into these new  properties 
by considering the effective Hamiltonian
\begin{equation}
	\mathcal{H}_{\sf R2U1}  \sim U \sum_\alpha (\partial_\beta  {E}_{\alpha \beta})^2 + \delta_1  (\Tr {\bm{E}})^2 +\delta_2  \sum_{\alpha < \beta} (E_{\alpha\beta} - E_{\beta\alpha})^2, 
\end{equation}
where \mbox{$U \sim J$} and \mbox{$\delta_{1,2}\sim D$}. 
In the temperature regime $T < D$, we find an emergent Gauss's law
\begin{equation}
	 \partial_\beta  {E}_{\alpha \beta}^\text{(sym.)} = 0
\end{equation}
in which ${E}_{\alpha \beta}^\text{(sym.)}$ is a rank-2 tensor, with only 
symmetric, traceless components.
Charge is now defined through a vector 
\begin{eqnarray}
	\rho_\alpha = \partial_\beta  {E}_{\alpha \beta}^\text{(sym.)} 
\end{eqnarray}
and subject to an extended set of conservation laws, 
which comprise conservation of charge
\begin{align} 
 	& \int d v\ \vec{\rho}=0 \; ,
 \end{align}
conservation of moment of inertia
\begin{align} 
  	& \int d v\ \vec{x} \times \vec{\rho}=-\int d v\ \epsilon_{\alpha \beta \gamma} E_{\beta \gamma}=0  \; ,
\end{align}
and conservation of dipole moment
\begin{align} 
  	& \int d v\ \vec{x} \cdot \vec{\rho}=-\int d v\ E_{\alpha \alpha   }=0 \; .
 \end{align}
Following \cite{pretko17a,pretko17b}, we refer to phase with these properties 
as a {\it rank--2} $U(1)$ spin liquid.  
The Gauss's law Eq.~(\ref{eq:Gausslaw}) also has important consequences for 
correlations, leading to pinch points in $S({\bf q})$ with a characterstic 4--fold structure~\cite{prem18a,Yan20a}.
}
 
%The ground states } form a magnetically disordered submanifold, described by a zero-divergence 
%\textit{tensor} field, which earned it the name of \textit{rank-2} spin liquid \cite{pretko17a,pretko17b}. 
%%
%The passages from paramagnet to rank-1 and then rank-2 spin liquids are both crossovers, 
%i.e. there is no spontaneous broken symmetry and no singularity from infinite temperature down 
%to $T_c$, as illustrated in Fig.~\ref{fig:ChChiT}. 
%%
%Such crossovers between distinct spin liquids have also been observed in the well-known 
%XXZ pyrochlore model~\cite{Taillefumier17a} and recently in a pyrochlore ferromagnet 
%with DM interactions \cite{lozano24a}.

%%%%%%%%%%%%%%%%%%%%%%%%%%%%%%%%%%%%%%%%%%%%%%%%%%%%%%%%%%
%%%%%%%%%%%%%%%%%%%%%%%%%%%%%%%%%%%%%%%%%%%%%%%%%%%%%%%%%%

%[Should define electric field, so we can say what it means for this to ``fluctuate''].

%%%%%%%%%%%%%%%%%%%%%%%%%%%%%%%%%%%%%%%%%%%%%%%%%%%%%%%%%%
\subsection{Undetermined order at low temperatures}
%%%%%%%%%%%%%%%%%%%%%%%%%%%%%%%%%%%%%%%%%%%%%%%%%%%%%%%%%%

\changes{
Finally, %\textcolor{orange}{for Heisenberg spins instead of the large-N limit spins,} 
when further decreasing the temperature for $D<0$, the rank-2 spin liquid 
was found to order at a temperature \mbox{$T_c \sim 10^{-3}\ J$}~\cite{Yan20a}.
However the nature of this order could not be determined from the   
analysis of Monte Carlo (MC) simulation carried out in Ref.~\cite{Yan20a}, 
because poor thermalisation made it impossible 
to carry out finite--size scaling of results for $T < T_c$.  
All that could be identified were potential Bragg peaks in the spin structure factor 
at finite $\mathbf{q}=W$ (corners of the Brillouin zone).
These results suggested the possibility of a spin order which broke
the symmetries of the lattice, in coexistence (or competition) 
with other low--energy states, but stopped short of identifying that order. 
}

%%%%%%%%%%%%%%%%%%%%%%%%%%%%%%%%%%%%%%%%%%%%%%%%%%%%%%%%%%
\subsection{Parameters studied in this Article}
%%%%%%%%%%%%%%%%%%%%%%%%%%%%%%%%%%%%%%%%%%%%%%%%%%%%%%%%%%

\changes{
It is the properties of the low--temperature ordered phase for $D < 0$, 
unidentified in Ref.~\cite{Yan20a}, which forms the main focus this Article.
To this end, all MC simulations 
% of Eq.~(\ref{eq:H}) %presented below 
were carried out for the parameter set  
\begin{eqnarray}
J=1 \; , \;  D=-0.0141 \; , 
\label{eq:parameters}
\end{eqnarray}
setting $k_B = 1$, and measuring temperature $T$ in units $10^{-3}\ J$.
In what follows, we review the status of MC simulations of the ordered phase, 
prior to AI input.
}\\

%%%%%%%%%%%%%%%%%%%%%%%%%%%%%%%%%%%%%%%%%%%%%%%%%%%%%%%%%%
\section{Simulations before AI input}
%%%%%%%%%%%%%%%%%%%%%%%%%%%%%%%%%%%%%%%%%%%%%%%%%%%%%%%%%%
\label{sec:MC}

\changes{
Monte Carlo (MC) simulations of the breathing--pyorchlore model Eq.~(\ref{eq:H}), 
for parameters Eq.~(\ref{eq:parameters}), were carried out following procedure used in Ref.~\cite{Yan20a}, 
for spins length $|S| = 1/2$, in clusters of up to $N=27\,648$ spins. 
The clusters considered were of cubic symmetry, with edges parallel to the (cubic) 
crystal axes.
These clusters contained $N= 16\ L^3$ spins, where $L$ is the linear dimension 
of the cluster, measured relative to a cubic unit cell which contains 16 spins, 
illustrated in Fig.~\ref{fig:quadpod}.
Lattice definitions are provided in Appendix~\ref{sec:Def}.
}

%%%%%%%%%%%%%%%%%%%%%%%%%%%%%%%%%%%%%%%%%%%%%%%%%%%%%%%%%%

\begin{figure}[ht]
\centering
\includegraphics[width=0.9\columnwidth]{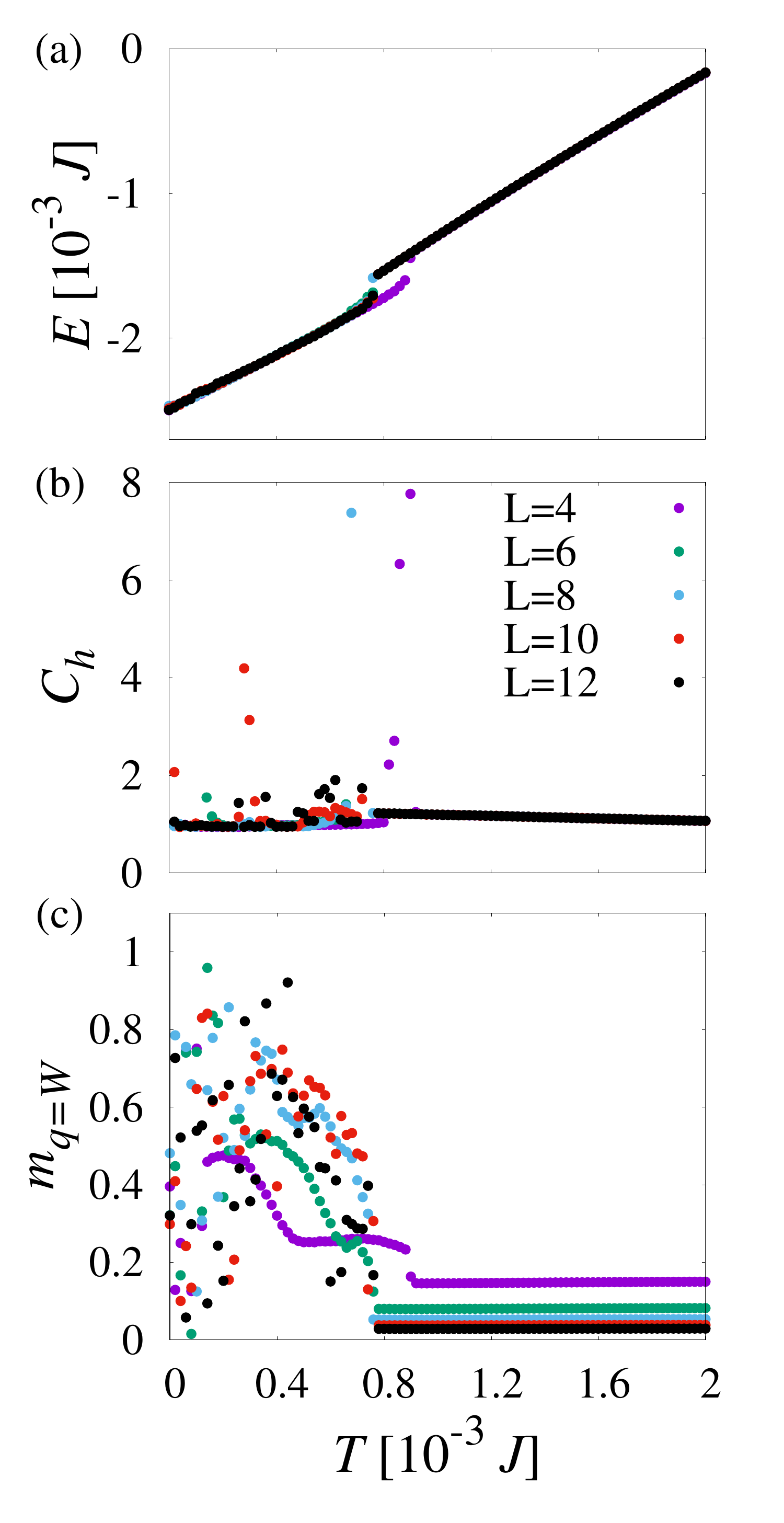}
\caption{Classical Monte Carlo (MC) simulation results for breathing pyrochlore model before input from 
machine learning, showing failure of simulations to equilibrate in the unidentified ordered 
phase for $T \lesssim 0.8\times10^{-3}\ J$. 
(a) Energy per site $E$.
(b) Heat capacity per site $C_h$
(c) Staggered magnetization $m_{\bf q}$ at the proposed ordering vector ${\bf q} = W$.    
Simulations were carried out for the model Eq.~(\ref{eq:H}), 
for a clusters with linear dimension $L \in \{4,6,8,10,12\}$, 
with parameters Eq.~(\ref{eq:parameters}), as described in Section~\ref{sec:MC}.
}
%\textbf{Monte Carlo simulations before AI input} show a phase transition at $T_c\approx 8.\,10^{-4} J$ in the (a) energy, (b) specific heat and (c) order parameter for wavevector $\mathbf{q}=W$ order (corners of the Brillouin zone) for system size $L\in \{4,6,8,10,12\}$. Simulations have been slowly annealed from high temperature during thermalisation. There are, however, noticeable finite-size effects and thermalisation issues below $T\sim 10^{-3} J$. Details of simulations are given in Sec.~\ref{sec:MC}. The energy in panel (a) has been shifted by $+J\,S^2=+1/4$ for convenience.
\label{fig:MC}
\end{figure}

Starting from a random spin configuration, the system is annealed from high temperature to a temperature $T$ during $10^6$ Monte Carlo steps, then thermalised at $T$ for another $10^6$ Monte Carlo steps, and finally data are collected for statistical averaging during $10^7$ MC steps. Each MC step is made of $N$ single-spin-flip updates via the rejectionless heatbath algorithm and five overrelaxation updates sweeping through the entire lattice. Overrelaxation is a micro-canonical update with spin rotation around the local molecular field for each spin; we include both $\pi$ rotations (the largest one) and random ones. Every 100 MC steps, there is parallel tempering between neighbouring temperatures (126 temperatures in parallel between $T=0$ and $0.0025\ J$). These simulation parameters are typically one order of magnitude longer in time and bigger in system size than what is usually necessary to completely characterise a typical phase transition in a frustrated magnet (see e.g. the simulation parameters in Ref.~[\onlinecite{Yan17a}]). Nonetheless, we see in Fig.~\ref{fig:MC} that while it is possible to spot the presence of long-range order, we cannot properly thermalise the magnetic order at very low temperatures. As a consequence it is unclear if the order with Bragg peaks at $\mathbf{q}=W$ is really (part of) the ground state and if it co-exists, or not, with other phases.

%%%%%%%%%%%%%%%%%%%%%%%%%%%%%%%%%
%. Fig. 3 - MC results before AI input
%%%%%%%%%%%%%%%%%%%%%%%%%%%%%%%%%

%%%%%%%%%%%%%%%%%%%%%%%%%%%%%%%%%

The reasons for this issue are multiple. On a fundamental level, our problem is the ordering mechanism arising upon cooling from a higher-rank gauge field. As was shown in Ref.~[\onlinecite{Yan20a}], the rank-2 gauge field itself comes out of a rank-1 gauge field with a broader phase space manifold, namely the Coulomb spin liquid of the Heisenberg anti-ferromagnet on pyrochlore. It is the DM term on $A-$tetrahedra in Eq.~\eqref{eq:H}, that selects the rank-2 gauge field. This means that the ordered phase we are investigating is separated from paramagnetic fluctuations by \textit{two} successive crossovers into more and more constrained configurational manifolds, see Fig.~\ref{fig:ChChiT}. Monte Carlo simulations are thus particularly constrained in phase space around $T_c$ and can easily be trapped in local free-energy minima.

This is where parallel tempering would usually help, by shuffling spin configurations across temperatures. But here the issue is not only that the transition temperature is far from paramagnetic fluctuations. The visible jumps in energy and order parameter in Fig.~\ref{fig:MC}(a,c) suggest a first-order transition. This strongly hinders the efficiency of parallel tempering, because the discontinuity in energy essentially prevents spin configurations from crossing the transition temperature. Hence one cannot rely on parallel tempering to help thermalise the ordered phase. In addition, the energy jump in Fig.~\ref{fig:MC}(a) is of the order of $10^{-4} J$. Such a tiny energy selection is consistent with the double crossover mentioned above but, keeping in mind that we are at proximity of a highly degenerate spin liquid, it also suggests a competition between multiple phases that are quasi-degenerate in free energy. Finally, the $\mathbf{q}=W$ Bragg peaks implies a large magnetic unit cell made of 32 sites, which is another complication in itself for the local heatbath algorithm.\\

Our point is that, even if not necessarily systematic, we expect thermalization issues to be relatively 
natural consequences of higher-rank gauge fields \cite{pretko17a,Nandkishore19a,pretko20a,Gromov24a}.  
Being governed by tensorial constraints and a multiplicity of conserved quantities \cite{pretko17a}, these exotic phases and dynamics are inherently complex, and it comes as to no-surprise for their ordering mechanisms to be unconventional. With these caveats in mind, our goal is to show how AI is able to help us, taking advantage of what it does best: extracting useful information out of noisy and incomplete data.

%%%%%%%%%%%%%%%%%%%%%%%%%%%%%%%%%%%%%%%%%%%%%%%%%%%%%%%%%%
\section{The machine learning algorithm}
%%%%%%%%%%%%%%%%%%%%%%%%%%%%%%%%%%%%%%%%%%%%%%%%%%%%%%%%%%
\label{sec:ML}

Our machine learning algorithm, the tensorial kernel support vector machine (TKSVM), has been developed in \lode{Refs.~[\onlinecite{Liu19a},\onlinecite{Greitemann19a},\onlinecite{Greitemann19b}]}. The inner working of the TKSVM algorithm are not necessary to understand the present work, as there are no algorithmic developments here, and we refer to \lode{Appendix~\ref{sec:TKSVM.overview}} for a concise introduction to the method. It is, however, important to know its input requirements and output. TKSVM takes as input Monte Carlo snapshots of the spin configurations (typically 500-1000 of them) between two different data sets. Since the Hamiltonian parameters are fixed, we will compare different temperatures, in particular above and below $T_c$. It also requires as input the definition of a (typically small) cluster of sites, specified by the user. TKSVM builds on these clusters on a tensorial basis up to a predefined rank. Rank one can be thought of as dipolar magnetic order, rank two as quadrupolar order etc. The stochastic quantities defined on this cluster are obtained by averaging spin configurations over the full lattice. This significantly reduces the data's dimension to a  dependence  on the size of the cluster. The output of TKSVM is the decision function used to separate data into sets with different characters. This comprises structure factors (in the term of a coefficient matrix) and a bias term. The bias, usually used to extract phase diagrams by TKSVM, plays no role in this work. The structure factors encode the order parameters squared, and can be interpreted by the user. In other words, in the (simplified) setting when the machine strips all the information differentiating two data sets down to a single scalar, the decision function encodes the local order parameter of Landau theory reflecting symmetry breaking between the two data sets. An additional strength of TKSVM is that even in the absence of order, it is able to measure local constraints hinting at a potential classical spin liquid candidate. The extent of the locality, for both the order parameter and the spin-liquid constraint, is limited by the cluster size. As an illustrative example we refer to Ref.~[\onlinecite{Greitemann19a}] where TKSVM successfully reproduced the phase diagram of the classical $XXZ$ model on the pyrochlore lattice computed by Taillefumier et al.~\cite{Taillefumier17a}: it found, and interpreted, the ordered in-plane ferromagnetic and nematic phases, but also identified and interpreted the crossovers between the high-temperature paramagnet and spin-ice, as well as the one with the Heisenberg spin liquid. Given this success and the tensorial nature of the higher-rank gauge fields, TKSVM is our method of choice to tackle our problem.

%-------------------------------------------------------------------------------------------------------------------------
%------------------------------------------------------------------------
\begin{figure}[t]
%\centering\includegraphics[scale=0.3]{2_n_cell_rank1_coll.png}
\centering\includegraphics[width=\columnwidth]{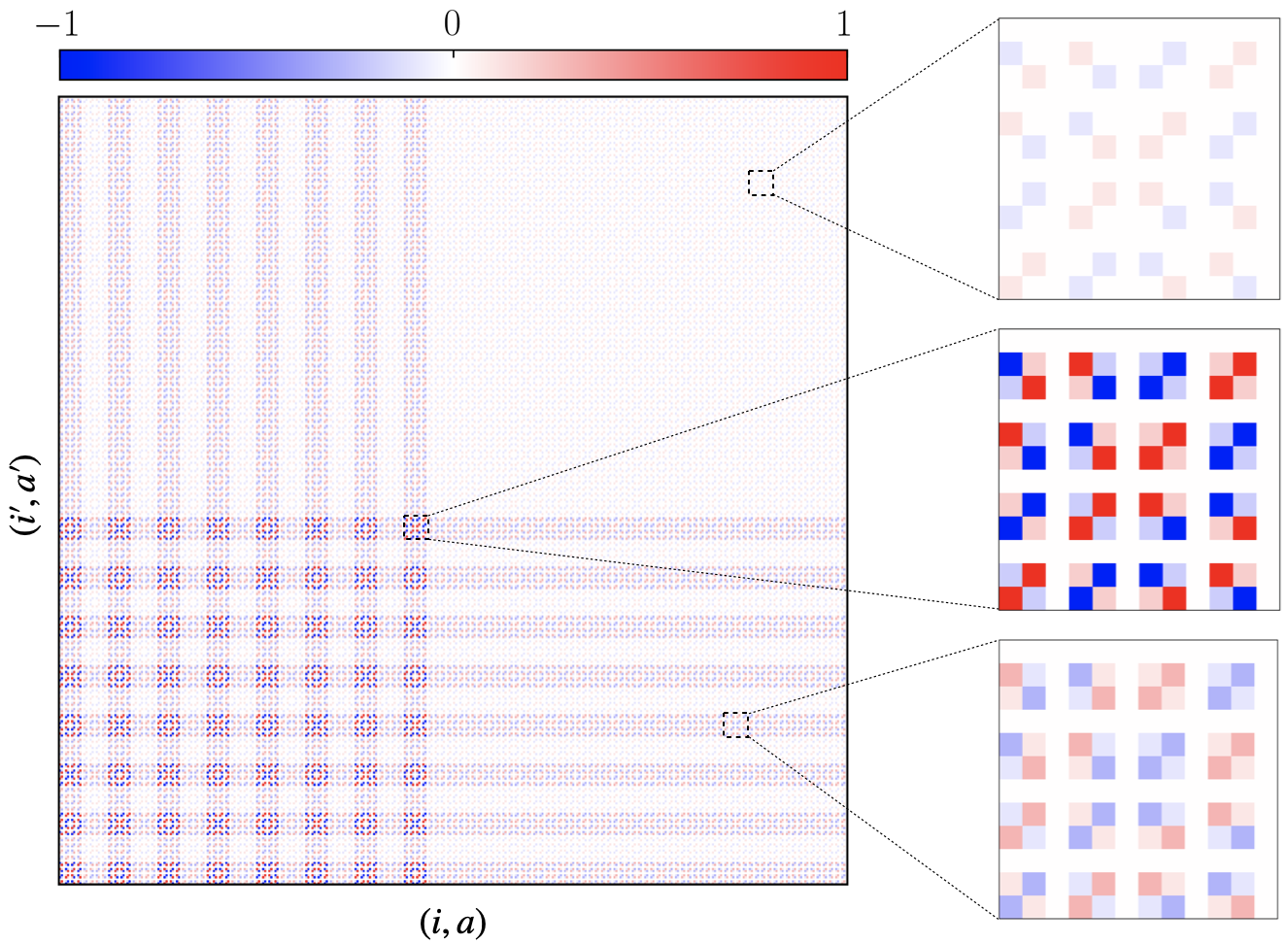}
\caption{
\textbf{Rank$-1$ coefficient matrix.} Each pixel represents the color-scaled weight of a contraction $\langle S_i^a\rangle \langle S_{i'}^{a'} \rangle$ of rank-1 features learned by the machine. The indices $i$ and $i'$ range from $0$ to $127$ labelling the spins within the $128$-site cluster and $a,a'$ label the spin components $x,y,z$. This $384 \times 384$ matrix is a graphical representation of the decision function for the rank$-1$ kernel [Eq.~(\ref{eq:C_munu})]. Magnification of three different spin-contraction blocks reveals their locally similar structure and different overall weight. Long-range magnetic order would have been revealed via a block pattern repeating itself across the entire matrix, which is absent here. \changes{The machine learned from spin configurations obtained by Monte Carlo simulations at $T= 4.~10^{-4}J$, well below the apparent phase transition at $T_c\approx 8.~10^{-4} J$ in Fig.~\ref{fig:MC}.}
}
\label{fig:Rank1Pattern}
\end{figure}
%------------------------------------------------------------------------

\subsection{Rank$-1$ results}
\label{sec:rank1}

Without prior knowledge of the phase, it is natural to start with the rank$-1$ kernel to probe potential magnetic orders \changes{well below the apparent phase transition of Fig.~\ref{fig:MC} at \mbox{$T_c\approx 8.~10^{-4} J$}}. Rank$-1$ means that we consider quantities which are linear combinations of the spin components of the cluster, or in other words dipolar forms of magnetic order. As the complexity of the feature vector grows linearly at rank$-1$ \cite{Greitemann19b}, we can use very large clusters consisting of multiple lattice unit cells. Provided a phase is purely magnetic and has a perfect translational symmetry, the rank$-1$ patterns learned with different cluster sizes should converge to a stable, regular, structure. The magnetic order parameter can then be inferred and justified a posteriori by measuring it in new Monte Carlo simulations.

However, in the unknown phase below $T_c$, we do not observe evidence of a stable rank$-1$ pattern even when using very large clusters up to 128 sites (8 cubic unit cells) in Fig.~\ref{fig:Rank1Pattern}. Instead, the learned patterns display sample-dependent irregular weights that are inconsistent with long-range dipolar order. This suggests that \changes{rank$-1$} magnetic orders do not reflect the correlations in the system fully, and we shall further inspect the data at rank$-2$ \changes{in order to confirm the presence, or not, of order below $T_c$}.

%-------------------------------------------------------------------------------------------------------------------------
\subsection{Rank$-2$ results}
\label{sec:rank2}

Rank$-2$ means that we consider quantities which are quadratic combinations of the spin components of the cluster, or in other words quadrupolar forms of magnetic order. The choice of the cluster at rank$-2$ is also guided by the lattice structure. Natural choices include a single $A$-tetrahedron, single $B$-tetrahedron, and FCC cubic unit cells of the breathing pyrochlore lattice consisting of $16$ spins, see Fig.~\ref{fig:quadpod}.
We found that considering the $A$-tetrahedron and $B$-tetrahedron separately reveals all the information of the phase. After obtaining results for these clusters separately, they need to be combined to understand the structure of the ground states. Alternatively, using a FCC unit cell as a cluster directly reveals the ground state structure, at the price of a more involved interpretation and some redundancy from the four $A$-tetrahedra contained in the cluster. For simplicity, we will present the information extracted from $A$- and $B$-tetrahedra successively.\\

%------------------------------------------------------------------------
\begin{figure}[t]
\centering
\includegraphics[width=\columnwidth]{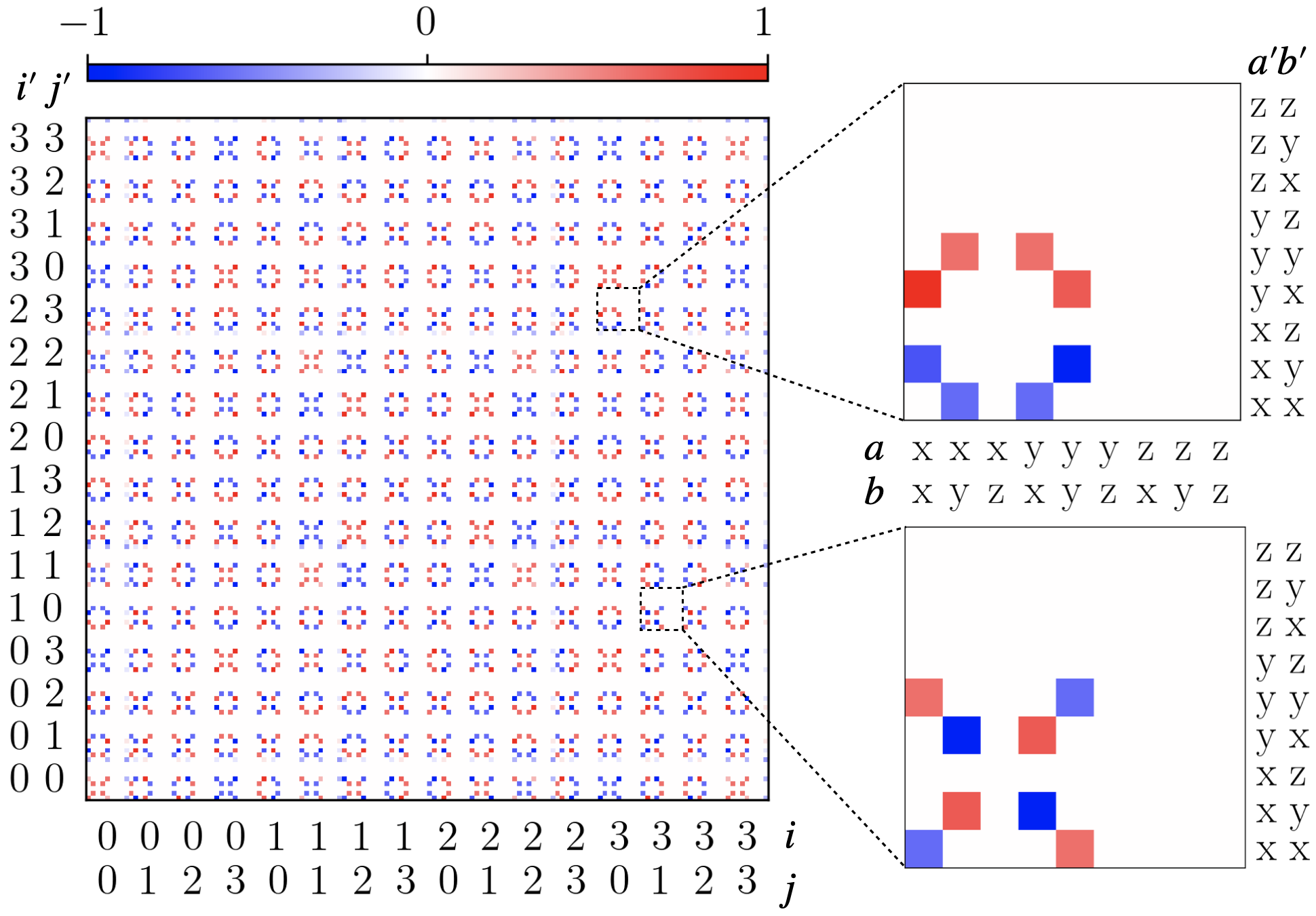}
\caption{
\textbf{Rank$-2$ coefficient matrix for $A$-tetrahedra}. Each pixel represents the color-scaled weight of a contraction $\langle S_i^a S_{j}^{b}\rangle \langle S_{i'}^{a'} S_{j'}^{b'} \rangle$ of rank-2 features learned by the machine. The $144 \times 144$ matrix is a graphical representation of the decision function $d_A^{xy}$, with spins confined to the $xy$ plane. Long-range quadrupolar order is revealed via the magnified block patterns repeating themselves across the entire matrix. The respective weights inside these block can be interpreted as an analytical expression of the rank$-2$ order parameter given in Eq.~(\ref{eq:dA}) (see Appendix ~\ref{sec:PatternInterpretation} for more details). \changes{The machine learned from spin configurations obtained by Monte Carlo simulations at $T= 4.~10^{-4}J$, well below the apparent phase transition at $T_c\approx 8.~10^{-4} J$ in Fig.~\ref{fig:MC}.}
}
\label{fig:AtetrPattern}
\end{figure}
%------------------------------------------------------------------------

%-------------------------------------------------------------------------------------------------------------------------
%\subsubsection{Order-by-disorder revealed by the $A$-tetrahedron}

For each of the four $A$-tetrahedra, the (sub-)decision function is
\begin{equation}\label{eq:dA}
d_A^s\sim (c_1^s)^2 + (c_2^s)^2,
\end{equation}
where $s \in \{xy, yz, zx\}$ labels the spin plane, which spontaneously breaks the spin permutation symmetry of the Hamiltonian in Eq.~\eqref{eq:H}. We stress that the spontaneous spin plane selection is uniform over all $A$-tetrahedra, and since every spin is part of one $A$-tetrahedron, the spin plane selection is \textit{global}. The precise order is thus defined by two effective order parameters $c_1$ and $c_2$ that are quadratic functions of spin components,
\begin{align}
c_1^{yz} &= \frac{1}{16} \big( (S_0^y - S_1^y + S_2^z - S_3^z)^2 + (S_0^z - S_1^z - S_2^y + S_3^y)^2 \big) \nonumber\\
c_1^{xz} &= \frac{1}{16} \big( (S_0^x + S_1^z - S_2^x - S_3^z)^2 + (S_0^z - S_1^x - S_2^z + S_3^x)^2 \big) \nonumber\\
c_1^{xy} &= \frac{1}{16} \big( (S_0^x + S_1^y - S_2^y - S_3^x)^2 + (S_0^y - S_1^x + S_2^x - S_3^y)^2 \big) \label{eq:definition-c1}
\end{align}
\begin{align}
c_2^{yz} &= \frac{2}{16} (S_0^y - S_1^y + S_2^z - S_3^z) (S_0^z - S_1^z - S_2^y + S_3^y) \nonumber\\
c_2^{xz} &= \frac{2}{16} (S_0^x + S_1^z - S_2^x - S_3^z) (S_0^z - S_1^x - S_2^z + S_3^x) \nonumber\\
c_2^{xy} &= \frac{2}{16} (S_0^x + S_1^y - S_2^y - S_3^x) (S_0^y - S_1^x + S_2^x - S_3^y).
\label{eq:definition-c2}
\end{align}

As TKSVM is conceived to learn the optimal order parameters, we can  reversely  infer maximally ordered spin configurations by maximizing $d_{A}^{s}$. These fully ordered states are potential ground states. Note that in order to facilitate the interpretability of TKSVM we shall not average over Monte Carlo samples in which different spin planes are spontaneously selected; in fact, it suffices to analyze all samples separately. With no loss of generality, we consider a state where the ordering develops in the spin $xy$ plane, whose corresponding TKSVM pattern (coefficient matrix) is illustrated in Fig.~\ref{fig:AtetrPattern}. The extraction of an analytical expression for the decision function from its graphical representation is discussed in Appendix ~\ref{sec:PatternInterpretation}.

%------------------------------------------------------------------------
\begin{figure}[t]
\centering
\subfloat[\label{subfig:AtetrConf}]{\includegraphics[width=0.4\columnwidth]{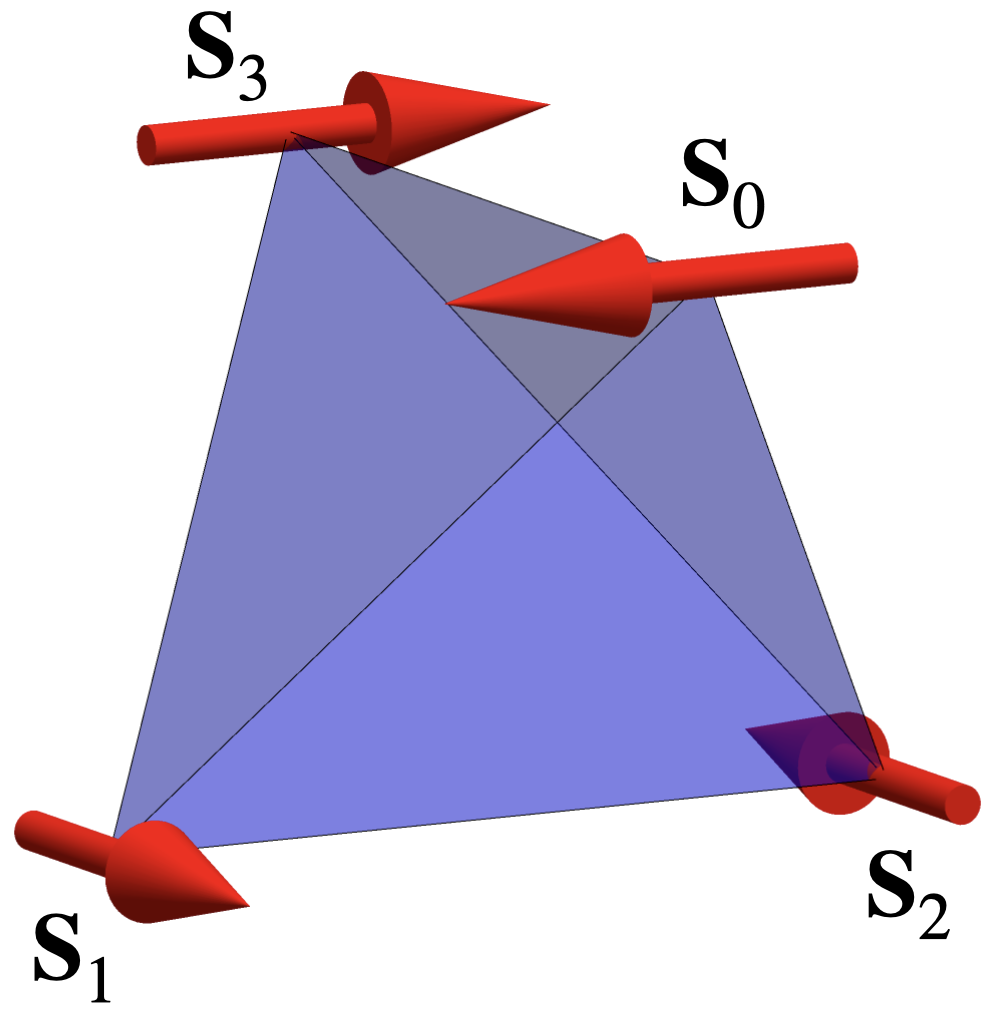}}
\quad\quad\quad
\subfloat[\label{subfig:c1c2E}]{\includegraphics[width=0.9\columnwidth]{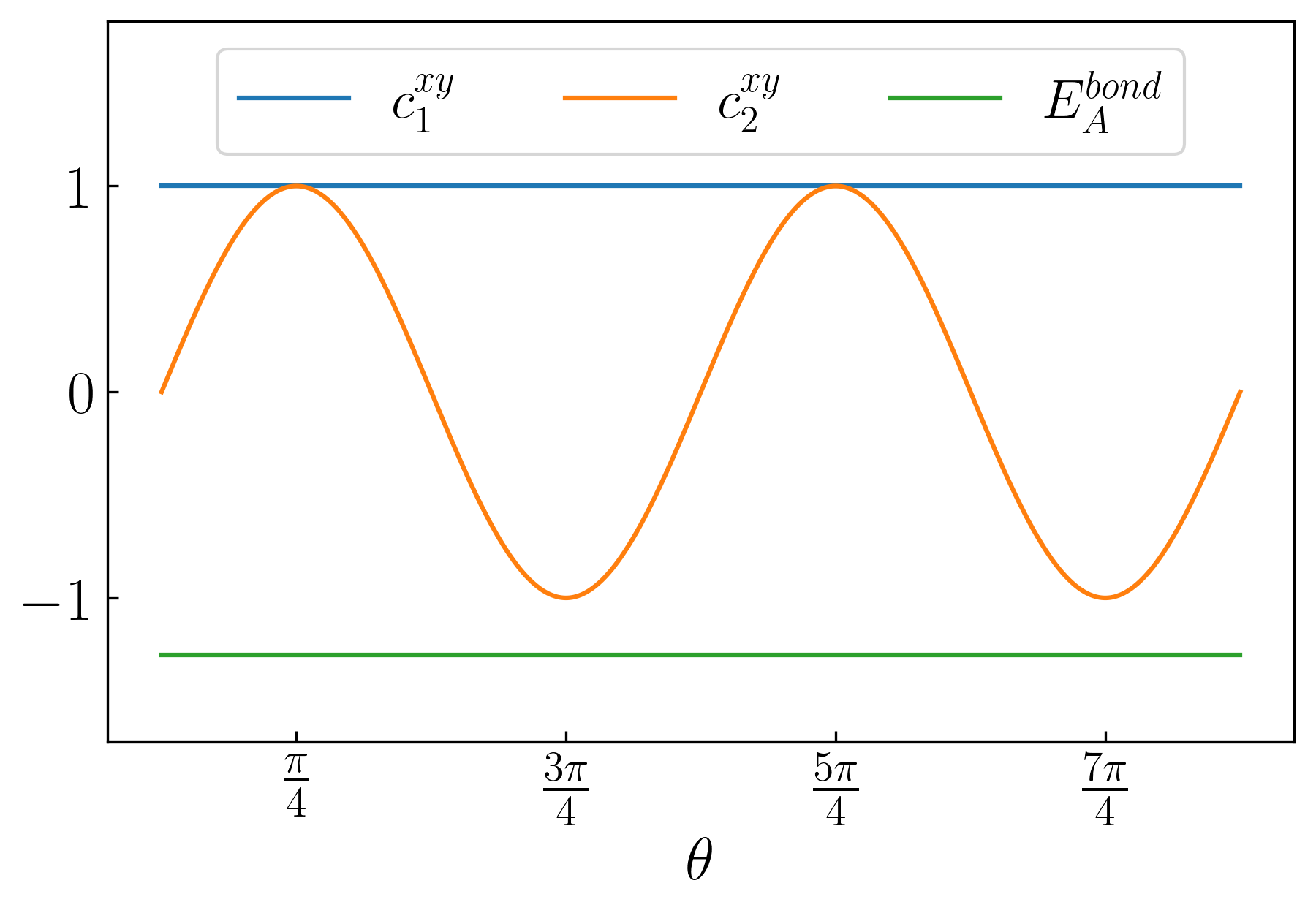}}
\caption{
\textbf{Order-by-disorder in $A-$tetrahedra.}
(a) Example of a state maximising both $c_1^{xy}$ and $c_2^{xy}$ with $\theta={7\pi}/{4}$ in Eq.~\eqref{eq:Aconfxy}. The bonds that lie inside the spin-plane ($0-3$ and $1-2$) have anti-parallel spins.
(b) The spin-plane-specific quantities $c_1^{xy}$ and $c_2^{xy}$ as well as the energy per bond, Eq.~\eqref{eq:EAbond}, with $J=1$ and $D=-2$] as functions of the parametrisation angle $\theta$. Note that in the decision function Eq.~\eqref{eq:dA} the term $c_2^{xy}$ appears squared, meaning that configurations with $c_2^{xy}=-1$ are also maxima of the decision function.
}
\label{fig:AtetrConf_c1c2E}
\end{figure}
%------------------------------------------------------------------------

%------------------------------------------------------------------------
\begin{table*}[t]
\renewcommand*{\arraystretch}{1.3}
\centering
\begin{tabular}{ c | c | c c c c c c | c }
\minibox[c]{Spatial \\ layer} & \minibox[c]{Maximal \\ weight} & $(\vekk{S}_0\cdot\vekk{S}_1)$  
         		& $(\vekk{S}_0\cdot\vekk{S}_2)$  & $(\vekk{S}_0\cdot\vekk{S}_3)$  & 
        	 	$(\vekk{S}_1\cdot\vekk{S}_2)$  & $(\vekk{S}_1\cdot\vekk{S}_3)$  & $(\vekk{S}_2\cdot\vekk{S}_3)$ 
	 		& \minibox[c]{missing \\ config.} \\
	 	\hline
		 $yz$ & $w_1 = 1 $ & -1 & $\pm 1$ & $\mp 1$ & $\mp 1$ & $\pm 1$ & -1  & $\ P(\vekk{\Lambda} _E^{\star1})=0$\\ 
		 $xz$ & $w_2 = 1 $ & $\pm 1$ & -1 & $\mp 1$ & $\mp 1$ & -1 & $\pm 1$  & $\ P(\vekk{\Lambda}_E^{\star2})=0$\\
		 $xy$ & $w_3 = 1$ & $\pm 1$ & $\mp 1$ & -1 & -1 & $\mp 1$ & $\pm 1$  & $\ P(\vekk{\Lambda}_E^{\star3})=0$\\
\end{tabular}
\caption{
\textbf{Collinear ground states on $B$-tetrahedra.} In each case there are two possible solutions, reflecting the $\mathbb{Z}_2$ symmetry. The leftmost column describes which spatial layer possesses the $\mathbb{Z}_2$ symmetry, while the rightmost column indicates which of the three possible configurations of the $B$-tetrahedra, defined in Eq.(\ref{eq:LambdaE}), is not allowed.}
\label{table:Bconf}
\end{table*}
%------------------------------------------------------------------------

%------------------------------------------------------------------------
\begin{figure*}[!ht]
\subfloat[\label{fig:BtetrPattern}]{\includegraphics[width=1.35\columnwidth]{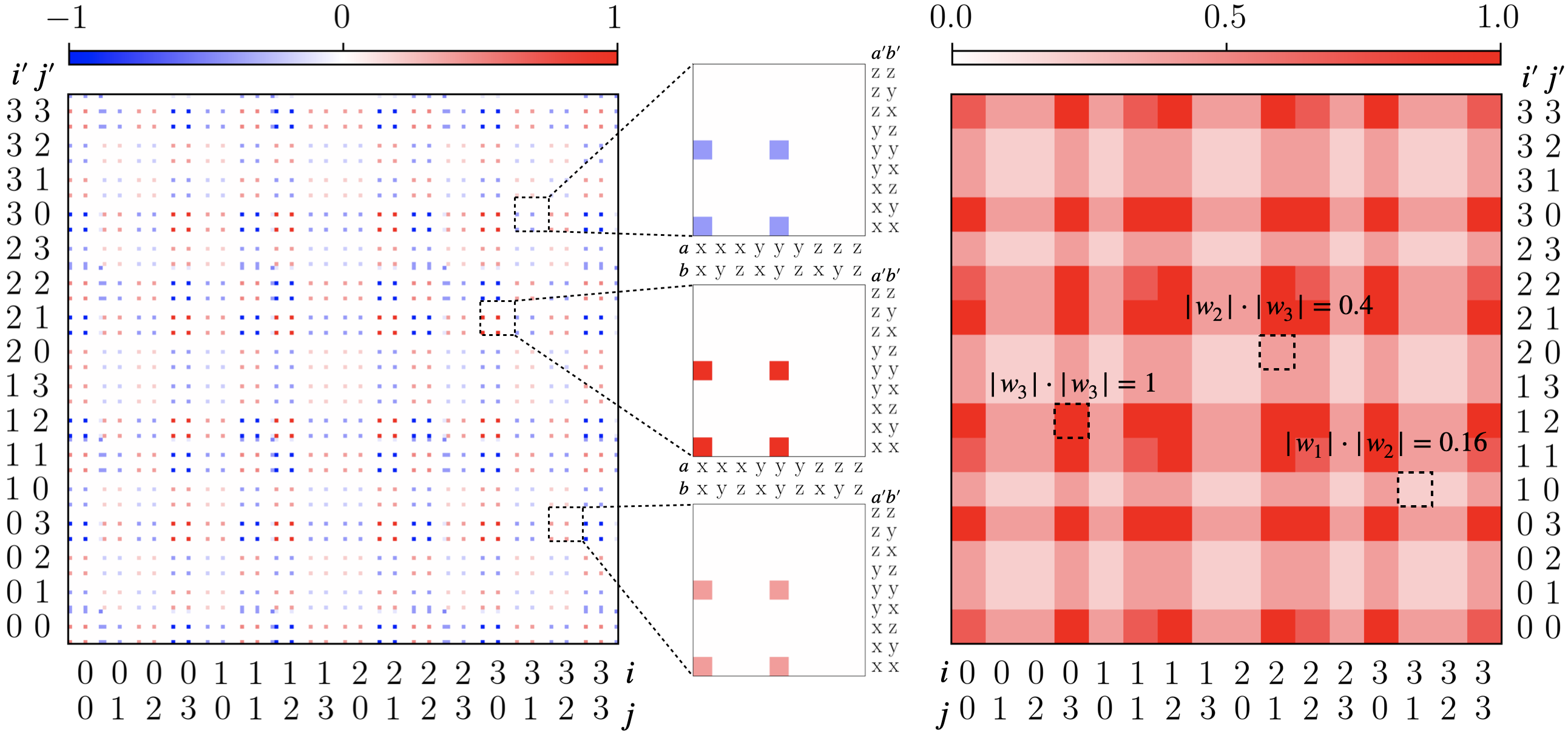}}\quad
\subfloat[\label{fig:quadpod_c2dist}]{\includegraphics[width=0.6\columnwidth]{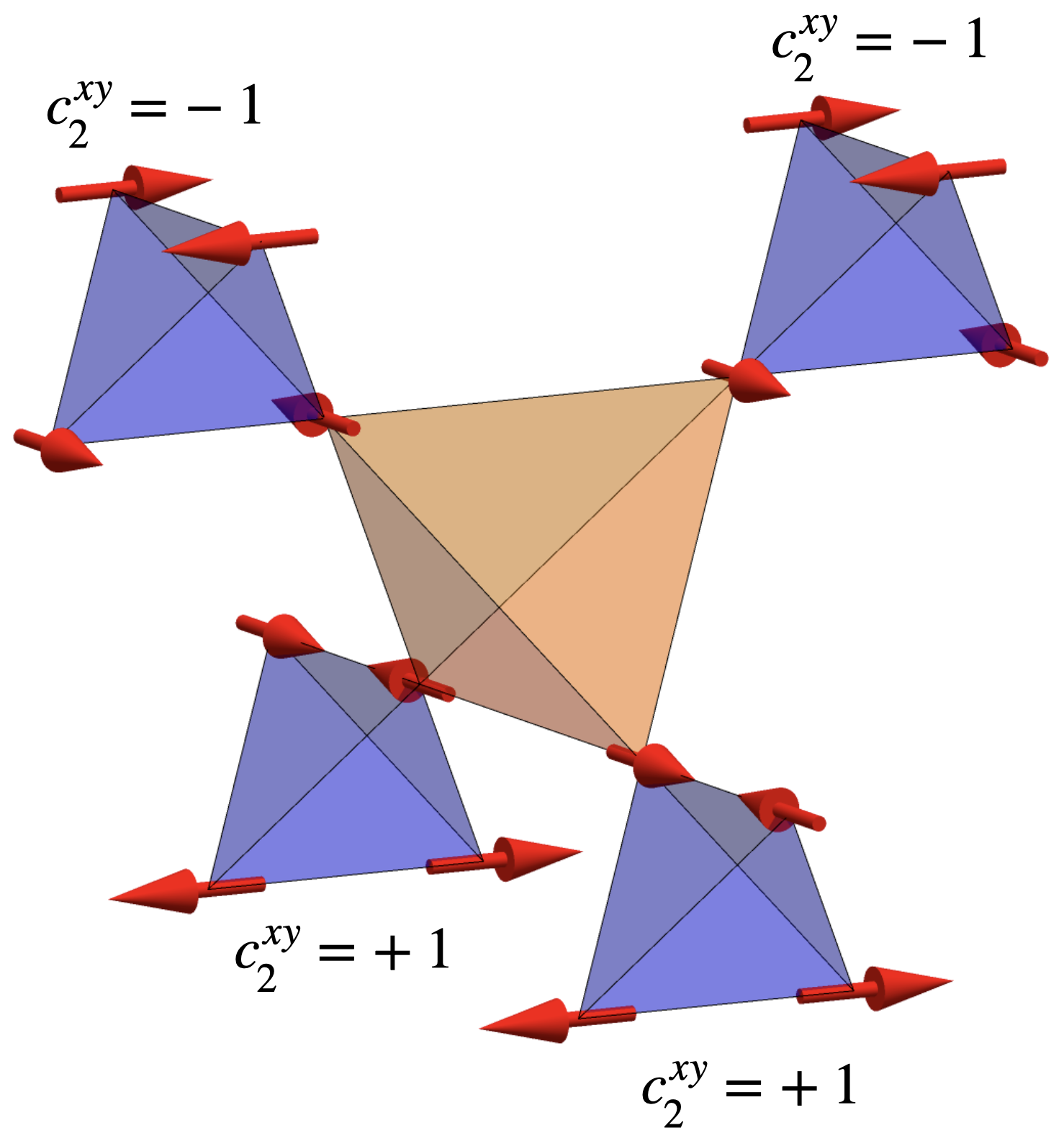}}
%\centering\includegraphics[scale=0.33]{tetrahedronB_coll.png}
\caption{
\textbf{(a) Rank$-2$ coefficient matrix for $B$-tetrahedra} as a graphical representation of the decision function $d_B$. Left: Each pixel represents the color-scaled weight of a contraction $\langle S_i^a S_{j}^{b}\rangle \langle S_{i'}^{a'} S_{j'}^{b'} \rangle$ of rank-2 features learned by the machine. All blocks have the same structure but different weights. Global spin plane selection confines all spins to the $xy$ plane. Right: Absolute weights of the coarse grained pattern. In this specific instance, the weights of Eq.~\eqref{eq:dBorig} are determined as $w_1=-w_2=0.4$ and $w_3=1$.
\textbf{(b) Ground-state configuration} with ordering in the $xy$ plane. The value of $c_2^{xy}$ is uniform within a $A$-tetrahedron layer, but alternates over different layers (cf. the upper two blue tetrahedra vs the lower two). In this example, the spin ordering plane coincides with the spatial plane.
}
\end{figure*}
%------------------------------------------------------------------------

\changes{
The form of the decision function, Eq.~\eqref{eq:dA}, suggests $c_1^s$ and $c_2^s$ 
are independent order parameters.
It follows that we can maximize $c_1^s$ and $c_2^s$ separately, and then check 
the consistency of their solutions.
(We return to this point Section~\ref{sec:Td}, where  $c_1^s$ and $c_2^s$ 
are linked explicitly to the symmetries of a tetrahedron).   
}
First, maximizing $(c_1^{xy})^2$ (i.e. solving $(c_1^{xy})^2 = 1$) leads to a manifold 
of spin configurations parametrized by an angle~\mbox{$\theta \in [0, 2\pi]$},
\begin{eqnarray}
\vekk{S}_0 = \begin{pmatrix}\cos{\theta}\\ \sin{\theta} \\0 \end{pmatrix}\quad
\vekk{S}_1 = \begin{pmatrix}\cos{\theta + \f\pi 2}\\ \sin{\theta + \f\pi 2} \\0 \end{pmatrix}\nonumber\\
\vekk{S}_2 = \begin{pmatrix}\cos{\theta - \f\pi 2}\\ \sin{\theta - \f\pi 2} \\0 \end{pmatrix}\quad
\vekk{S}_3 = \begin{pmatrix}\cos{\theta + \pi}\\ \sin{\theta + \pi} \\0 \end{pmatrix},
\label{eq:Aconfxy}
\end{eqnarray}
as illustrated in Fig.~\ref{subfig:AtetrConf}. Maximizing $(c_2^{xy})^2$ falls into the same structure of Eq.~\eqref{eq:Aconfxy}. Its evolution as a function of $\theta$ is plotted in Fig.~\ref{subfig:c1c2E} and shows that $(c_2^{xy})^2$ is maximised for four discrete values only:
\begin{eqnarray}
\theta\in\left\{\f{\pi}{4},\f{3\pi}{4},\f{5\pi}{4},\f{7\pi}{4}\right\}.
\label{eq:spThetas}
\end{eqnarray}

The $U(1)$ manifold of Eq.~\eqref{eq:Aconfxy} belongs to the ground state of the traditional pyrochlore anti-ferromagnet with negative DM interactions on \textit{all} tetrahedra \cite{canals08a,chern10b,noculak23a}, i.e. without breathing anisotropy. Indeed, applying the spin configuration of Eq.~\eqref{eq:Aconfxy} to the Hamiltonian of Eq.~\eqref{eq:H}, we recover its ground-state energy per bond \cite{canals08a}
\begin{equation}
E_A^{\rm bond} = \frac{1}{6}(-2J + 2\sqrt{2} D).
\label{eq:EAbond}
\end{equation}
Since the ground-state energy of Eq.~\eqref{eq:EAbond} is independent of $\theta$, any selection of specific $\theta$ values is necessarily due to thermal order by disorder (ObD), i.e. the selection is entropic rather than energetic.

In the standard model without breathing anisotropy, the ObD mechanism selects the $\mathbf{q}=0$ \changes{$\Gamma_5$} long-range order \cite{canals08a,chern10b,noculak23a}. \changes{Imposing $\theta=\f{3\pi}{4}$ (or $\theta=\f{7\pi}{4}$) on all tetrahedra corresponds to such a $\Gamma_5$ state, as} illustrated in Fig.~\ref{subfig:AtetrConf}.

However, in our model where DM terms disappear on $B-$tetrahedra, the $B-$tetrahedra \changes{\textit{locally} bear the ground-state degeneracy of the pyrochlore antiferromagnet. This local enhancement of the ground state degeneracy implies that $B-$tetrahedra do not have to order into \changes{$\Gamma_5$} states anymore. As a consequence,} the four $A-$tetrahedra surrounding a $B-$tetrahedron \changes{are less constrained and are thus} not forced to be in the same state; they have the freedom to explore different parts of the U(1) ground state manifold \changes{of Eq.~\eqref{eq:Aconfxy}} independently from each other. This enhancement of the ground state manifold is the reason why the ground state of our model of Fig.~\ref{fig:GS} is not simply a \changes{$\Gamma_5$} state, as confirmed by the machine which finds two additional solutions on the $A-$tetrahedra, $\theta\in\left\{\f{\pi}{4},\f{5\pi}{4}\right\}$.

At this stage, the ordering mechanism is thus a two-step process. The first step is the selection of the $U(1)$ manifold, and second is the coalescence on special points of the manifold via thermal order-by-disorder. It is remarkable that the machine is able to extract both sets of solutions, by which we can infer the order-by-disorder phenomenon, out of noisy numerical data.\\

%-------------------------------------------------------------------------------------------------------------------------
%\subsubsection{Collinear configuration on the $B$-tetrahedron}

Now let us focus on the other type of tetrahedra, with only anti-ferromagnetic couplings (no DM terms). The (sub-)decision function on the $B$-tetrahedra is identified as 
\begin{eqnarray}
d_B\sim \big[\ 
w_1\ (\mathbf{S}_0\cdot\mathbf{S}_1 + \mathbf{S}_2\cdot\mathbf{S}_3)
+w_2\ (\mathbf{S}_0\cdot\mathbf{S}_2 + \mathbf{S}_1\cdot\mathbf{S}_3)\nonumber\\
+w_3\ (\mathbf{S}_0\cdot\mathbf{S}_3 + \mathbf{S}_1\cdot\mathbf{S}_2)\big]^2.
\label{eq:dBorig}
\end{eqnarray}
The values of the weights $w_1, w_2, w_3$ can be inferred from the TKSVM pattern in Fig.~\ref{fig:BtetrPattern}, satisfying
\begin{align}\label{eq:Bweights}
	w_1 + w_2 + w_3 = 1, \quad \max w_i = 1,
\end{align} 
where the maximal $w_i$ is related to the ordering spin plane in $d_A^s$. Under this constraint, $d_B$ can be intuitively maximized if the four spins in a $B$-tetrahedron are {\it collinear}. The solutions are listed in Table~\ref{table:Bconf}, and Fig.~\ref{fig:BtetrPattern} shows an example of the coefficient matrix when the ordering is in the spin $xy$ plane with $w_3 = 1$.\\

Alternatively, the collinearity on the $B$-tetrahedra can also be derived from the constraints on the $A-$tetrahedra derived in the previous section. Without losing generality, we again take $d_A^{xy}$ as an example. The four solutions of $\left(c_2^{xy}\right)^2 = 1$ can be divided into two classes
\begin{align} 
c_2^{xy} = 1: \quad
&\vekk{S}_0 = \frac{1}{\sqrt{2}}\begin{pmatrix} 1 \\ 1 \\0 \end{pmatrix}\quad
\vekk{S}_1 = \frac{1}{\sqrt{2}}\begin{pmatrix} -1 \\ 1 \\0 \end{pmatrix}\nonumber\\
&\vekk{S}_2 = \frac{1}{\sqrt{2}}\begin{pmatrix} 1 \\ -1 \\0 \end{pmatrix}\quad
\vekk{S}_3 = \frac{1}{\sqrt{2}}\begin{pmatrix} -1 \\ -1\\0 \end{pmatrix},
\label{eq:special_c2_1}
\end{align}
\begin{align} 
c_2^{xy} = -1: \quad
&\vekk{S}_0 = \frac{1}{\sqrt{2}}\begin{pmatrix} 1 \\ -1 \\0 \end{pmatrix}\quad
\vekk{S}_1 = \frac{1}{\sqrt{2}}\begin{pmatrix} 1 \\ 1 \\0 \end{pmatrix}\nonumber\\
&\vekk{S}_2 = \frac{1}{\sqrt{2}}\begin{pmatrix} -1 \\ -1 \\0 \end{pmatrix}\quad
\vekk{S}_3 = \frac{1}{\sqrt{2}}\begin{pmatrix} -1 \\ 1\\0 \end{pmatrix},
\label{eq:special_c2_2}
\end{align}
up to a global sign flip which preserves the value of $c_2^{xy}$. Eqs.~\eqref{eq:special_c2_1} and \eqref{eq:special_c2_2} correspond to $\theta\in\left\{\f{\pi}{4},\f{5\pi}{4}\right\}$ and $\left\{\f{3\pi}{4},\f{7\pi}{4}\right\}$ respectively. Measurements of this machine-learned quantity show that $c_2^{xy} = \pm 1$ {\it alternates} through the $z$ direction, as shown in Fig.~\ref{fig:quadpod_c2dist}. As a $B$-tetrahedron shares spins with four $A$-tetrahedra, it has to take two collinear spins from Eq.~\eqref{eq:special_c2_1} and then two spins with the same collinearity from Eq.~\eqref{eq:special_c2_2}. Otherwise, it cannot satisfy the staggered distribution $c_2^{xy} = \pm 1$. This long-range order breaks spin-rotation, spin-permutation and translational symmetry.

%This collinearity adds another key element for inferring an emergent $\mathbb{Z}_2$ planar symmetry which will be discussed in Sec.~\ref{sec:Z2symmetry}.

%------------------------------------------------------------------------
%\begin{figure}[!t]
%\centering
%\includegraphics[width=0.4\textwidth]{fig_quadpod_c2dist.pdf}
%\caption{\textbf{Ground-state configuration} with ordering in the $xy$ plane. The value of $c_2^{xy}$ is uniform within a $A$-tetrahedron layer, but alternates over different layers (cf. the upper two blue tetrahedra vs the lower two). In this example, the spin ordering plane coincides with the spatial plane.}
%\label{fig:quadpod_c2dist}
%\end{figure}
%------------------------------------------------------------------------

%%%%%%%%%%%%%%%%%%%%%%%%%%%%%%%%%%%%%%%%%%%%%%%%%%%%%%%%%%
\section{Simulations after AI input}
%%%%%%%%%%%%%%%%%%%%%%%%%%%%%%%%%%%%%%%%%%%%%%%%%%%%%%%%%%
\label{sec:MC2}

\begin{figure}[t]
\centering
\includegraphics[width=0.9\columnwidth]{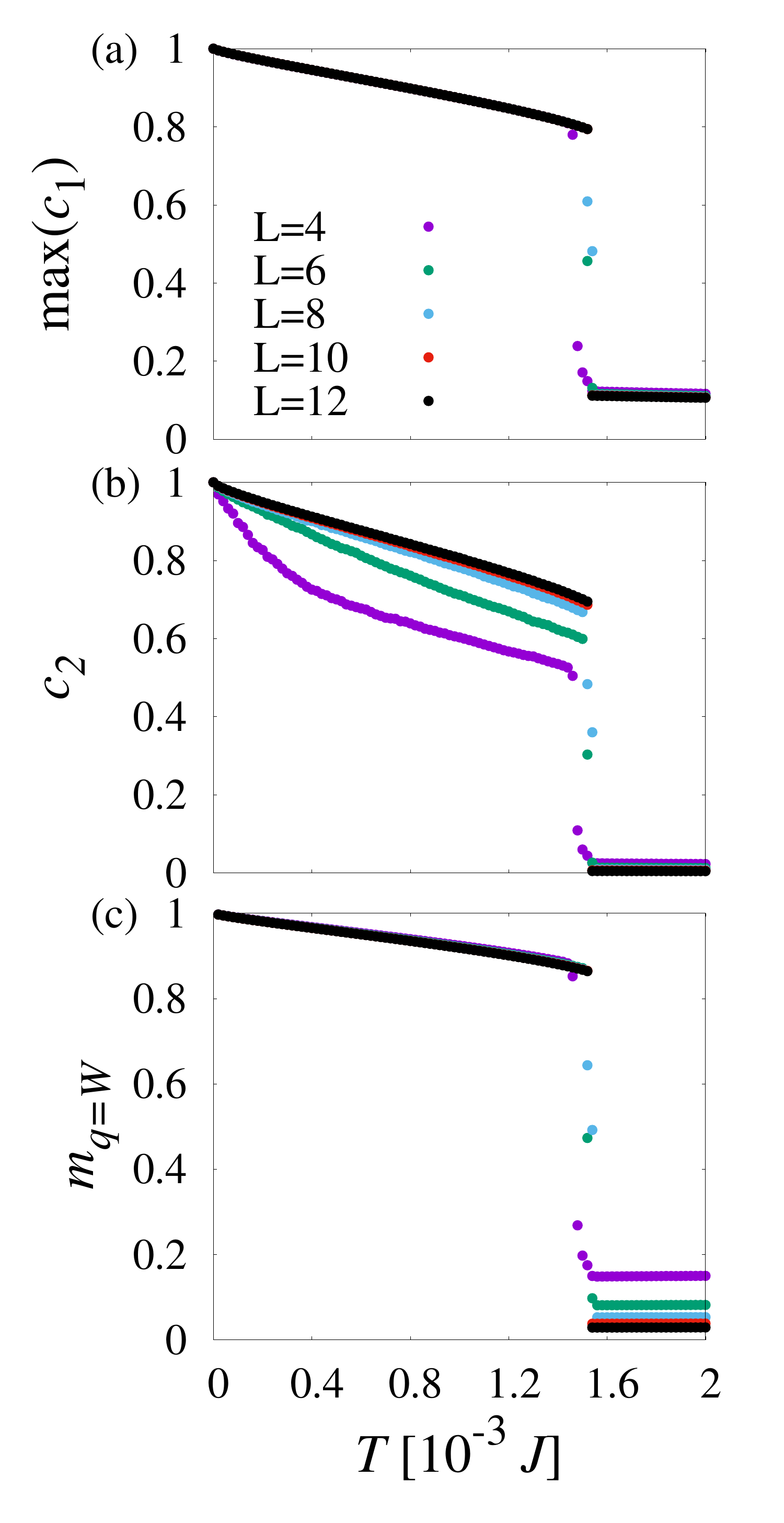}
\caption{Classical Monte Carlo (MC) simulation results for breathing pyrochlore model after input from 
machine learning, showing good equilibriation in the ordered phase for $T \lesssim 1.5\times10^{-3}\ J$. 
(a) Rank--one order parameter $c_1$, corresponding to dipolar order
(b) Rank--two order parameter $c_2$, corresponding to quadrupolar order.
(c) Staggered magnetization $m_{\bf q}$ at wave vector ${\bf q} = W$.  
Simulations were carried out for the model Eq.~(\ref{eq:H}), 
for a clusters with linear dimension $L \in \{4,6,8,10,12\}$, 
with parameters Eq.~(\ref{eq:parameters}).  % $D=-0.0141\ J$.
Further details, including the definitions of the order parameters $c_1$ and $c_2$ 
can be found in Section~\ref{sec:MC2}.
}
%\textbf{Monte Carlo simulations after machine input} shows a phase transition at $T_c\approx 1.5\, 10^{-3} J$ in all three order parameters: (a) the average of the maximum value of $c_1$, (b) $c_2$ and (c) the order parameter for wavevector $\mathbf{q}=W$ order (corners of the Brillouin zone) for system size $L\in \{4,6,8,10,12\}$. Simulations have been quenched into the state of Fig.~\ref{fig:GS} for each temperature.
\label{fig:MC2}
\end{figure}

Equipped with the new insights on the structure of the ground state, we return to Monte Carlo simulations. 
This time instead of slowly annealing from high temperature, we initialize the system by quenching 
into the configuration of Fig.~\ref{fig:GS} at temperature $T$, followed by $10^6$ MC steps of thermalisation, 
and $10^7$ MC steps for measurements. 
Paving the lattice with the ground state found by the machine requires alternating along the $z$ direction, 
the $xy-$layers of spin configurations as in Fig.~\ref{fig:quadpod_c2dist} with their time-reversal symmetric. 
This naturally forms a 32-site magnetic unit cell, see Fig.~\ref{fig:GS}. 
We have again 126 temperatures equally spaced between 0 and $0.0025 J$, and use heatbath, 
parallel tempering and over-relaxation algorithms. 
The results are shown in Fig.~\ref{fig:MC2}, computed for the same physical parameters mentioned in Sec.~\ref{sec:MC}.

%%%%%%%%%%%%%%%%%%%%%%%%%%%%%%%%%
%. Fig. 4 - MC results after AI input
%%%%%%%%%%%%%%%%%%%%%%%%%%%%%%%%%

%%%%%%%%%%%%%%%%%%%%%%%%%%%%%%%%%

These new MC simulations converge nicely and confirm the stability of the ground state found 
by the machine. 
The transition is now violently first order, whose hysteresis explains the shift of the transition temperature $T_c$ between $0.8\,10^{-3} J$ in Fig.~\ref{fig:MC} and $1.5\,10^{-3} J$ in Fig.~\ref{fig:MC2}. 
Quenched simulations in the latter case provide an upper bound of $T_c$, while slow annealing has more difficulty in finding the ordered phase and provides a lower bound to $T_c$. The $c_1$ and $c_2$ order parameters correctly describe the ground state, with a noticeably stronger finite-size dependence for the latter; a common consequence of the order-by-disorder mechanism \cite{Yan17a}. Finally, the order parameter $m_{q=W}$ now saturates at $T=0$, which means there is a priori no co-existence of other phases.

%%%%%%%%%%%%%%%%%%%%%%%%%%%%%%%%%%%%%%%%%%%%%%%%%%%%%%%%%%
\section{Emergent $\mathbb{Z}_2$ planar symmetry}
%%%%%%%%%%%%%%%%%%%%%%%%%%%%%%%%%%%%%%%%%%%%%%%%%%%%%%%%%%
\label{sec:Z2symmetry}

We now put the machine-learned quantities together and discuss an emergent planar-flip symmetry, which will also resolve the origin of the irregular weights in Fig.~\ref{fig:Rank1Pattern}. For simplicity, we continue to consider the ground state of Fig.~\ref{fig:GS} as an example, where the order is developed in the $xy$ plane and the planar symmetry acts on spatial $xy$ planes. In general these two planes do not need to coincide, but this does not affect our discussion: the spin-order plane is manifest from the $c_1$ and $c_2$ parameters, while the direction of the spatial planar-flip symmetry can be known from the largest weight in $d_B$.
 
Given the collinearity on the $B$-tetrahedron, the decision function Eq.~\eqref{eq:dBorig} reduces to
\begin{align}\label{eq:dB_reduced}
	d_B &\sim (-w_1 -w_2 + w_3)^2 = (w_3)^2.
\end{align}
Here we have used the solutions in Eqs.~\eqref{eq:special_c2_1} and~\eqref{eq:special_c2_2} (or equivalently the corresponding configuration in Table~\ref{table:Bconf}) and the weight relation in Eq.~\eqref{eq:Bweights}.
As in this example $\max w_i = w_3 = 1$, the relation Eq.~\eqref{eq:Bweights} reduces to $w_1 + w_2 = 0$.

\begin{figure}[!ht]
\centering
\subfloat[\label{subfig:planarFlip3D}]{\includegraphics[width=\columnwidth]{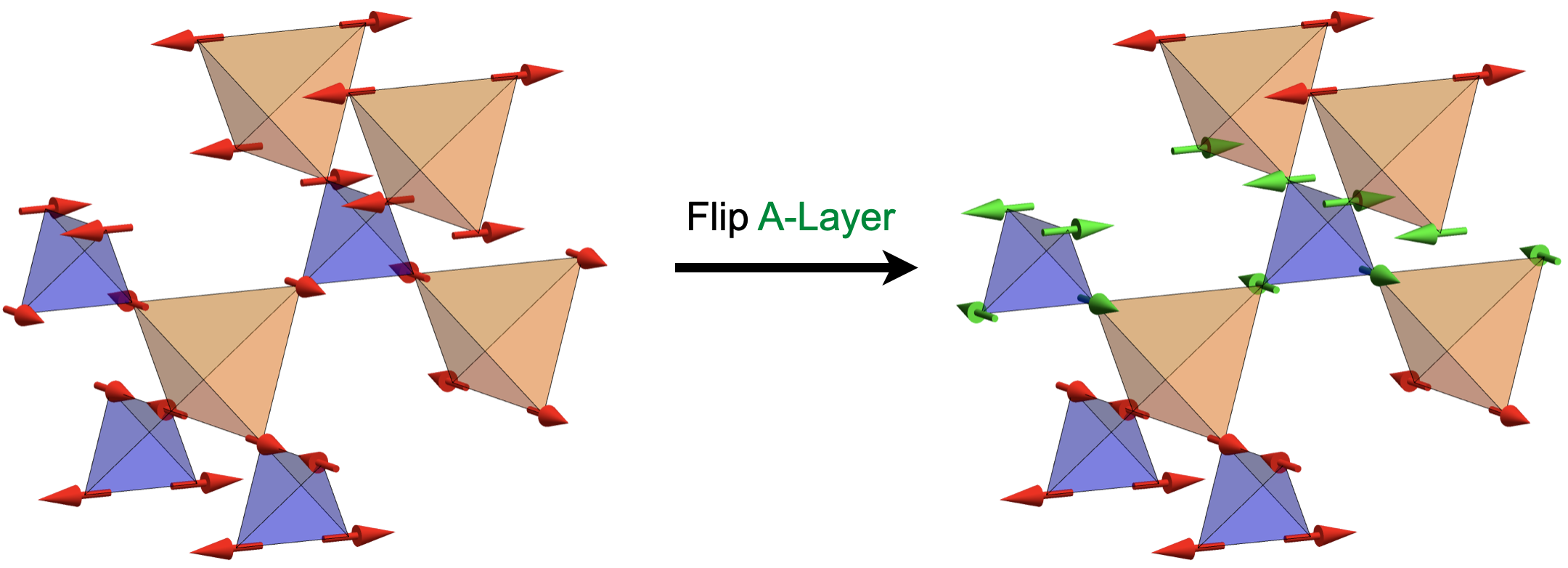}} \\ 
\subfloat[\label{subfig:planarFlip2D}]{\includegraphics[width=\columnwidth]{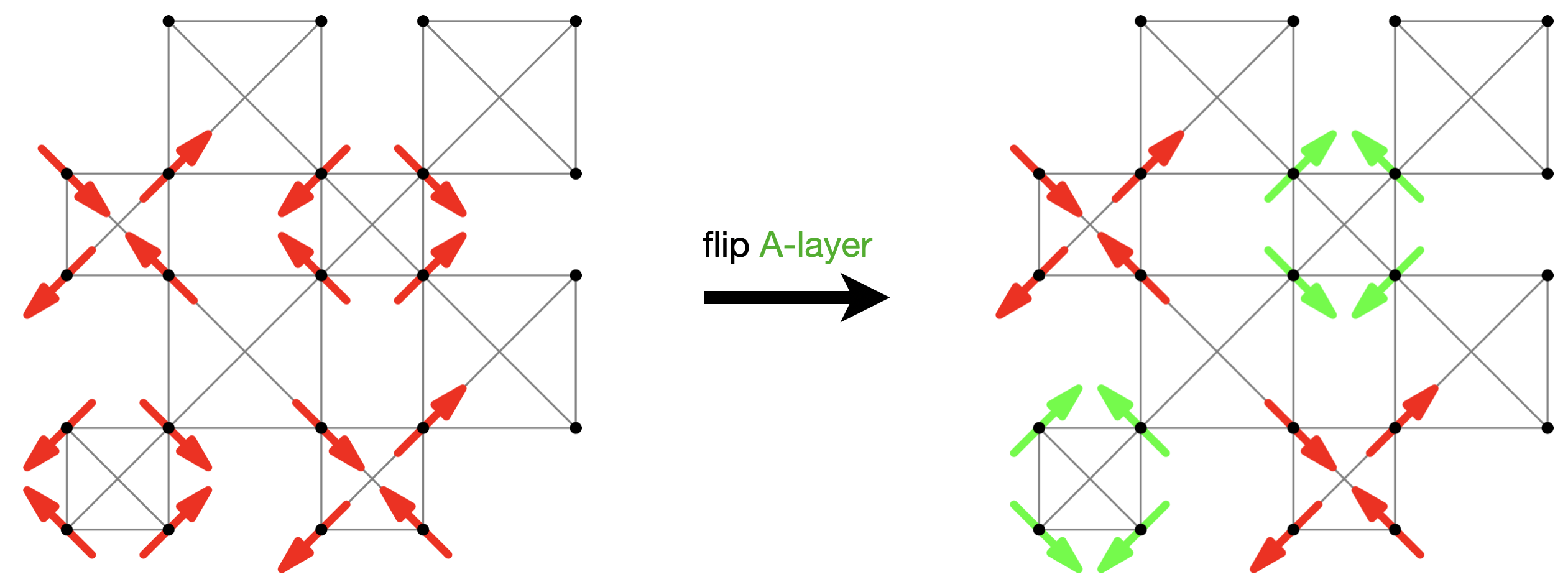}}
\caption{
\changes{Visual representation of planar} $\mathbb{Z}_2$ symmetry.
(a)~All $A$-tetrahedra within one layer are flipped, \changes{following Eq.~(\ref{eq:Z2.symmetry})}, 
thereby changing the value of $\vekk{\Lambda}_E$ [Eq.~(\ref{eq:bondop})] in the two 
neighbouring $B$-layers. 
On $B$-tetrahedra, only intra-layer bonds connecting two $A$-tetrahedra from the same layer 
are affected by the transformation. 
(b)~\changes{Equivalent transformation, viewed in a} projection onto the $xy$-plane; for simplicity not all spins are drawn. This example shows a case where the normal on the spin plane coincides with the normal on the layer-partitioning, but in general we can also construct ground states where the two normals do not coincide.
\changes{In all cases, the spins changes by the $\mathbb{Z}_2$ transformation are indicated in green.}
}
\label{fig:planarFlip}
\end{figure}

Eq.~\eqref{eq:dB_reduced} manifests a property of $d_B$ that it is invariant under flipping a specific pair of spins, which can be $(\mathbf{S}_0, \mathbf{S}_3)$ or $(\mathbf{S}_1, \mathbf{S}_2)$ in the current example.
Nevertheless, as spins in a $B$-tetrahedron belong to different $A$-tetrahedra, in order to preserve the value of the order parameters $c_1$ and $c_2$, one has to flip all spins in the two neighbouring $A$-tetrahedra. This procedure is then repeated to further $A$-tetrahedra neighbours, and closes only after flipping 
\changes{the spins in an entire layer $dA$ of $A$-sublattice tetrahedra 
\begin{eqnarray}
	{\bf S}_i  \to - {\bf S}_i \quad \forall \quad i \in \ dA
\label{eq:Z2.symmetry}
\end{eqnarray}
}
as illustrated in Fig.~\ref{fig:planarFlip}.

%------------------------------------------------------------------------
%------------------------------------------------------------------------

Namely, there is an {\it emergent} subsystem symmetry acting on individual ordering layers of $A$-tetrahedra (equivalently, two adjacent $B$-tetrahedron layers, while only the bottom or top half of each $B$-layer is transformed). As a consequence, the system ``hybridizes'' rank$-1$ and rank$-2$ orders. This is a rather unconventional emergent property for an ordered phase; such sub-extensive zero modes usually require to be artificially enforced via either a global or local symmetry of the system~\cite{Nandkishore19a,pretko20a,Gromov24a}.

Since this planar spin flip is sub-extensive -- the number of spins involved scales as $L^2$ -- we expect it to be dynamically robust. Pragmatically, let's assume that while cooling down the system, such a planar spin flip takes place. This is quite possible since it costs zero energy and the transition is violently first order, see Fig.~\ref{fig:MC2}; once the cubic symmetry is broken in favour of a given plane (here the $xy$ plane), two far-away layers of $A-$tetrahedra may likely order independently from each other. As a result, the long-range dipolar order with Bragg peaks at  $\mathbf{q}=W$ found in Fig.~\ref{fig:MC2} will be perturbed by multiple planar spin flips at random positions in the system. And since there is a vanishingly small probability to move $L^2$ spins coherently in the thermodynamic limit, such energetically degenerate spin configuration should remain stable over very long time scales. 

This mechanism explains the origin of the irregular weights in Fig.~\ref{fig:Rank1Pattern}, and probably played an important role in the difficulty to interpret previous Monte Carlo simulations, by hindering the ordering mechanism and hiding the magnetic dipolar order. As a result, the magnetic order will only be long-range within the plane.
The inter-plane ordering is quadrupolar and defined by $c_1$ and $c_2$ in Eq.~\eqref{eq:definition-c1} and Eq.~\eqref{eq:definition-c2}, akin to spin nematics~\cite{Mendels11a} but only in the spatial direction orthogonal to the magnetic layers. 
Moreover, the translational symmetry is preserved along this orthogonal direction, while the lattice rotation symmetry is spontaneously broken due to the emergent planar spin flips. 
This is reminiscent of the smectic phase in liquid crystal physics, which breaks spatial rotational symmetry but only partially breaks translational symmetries~\cite{DeGennesProst95}.
\changes{
Therefore, in addition to the in-plane magnetic orders, such states share interesting features of both spin nematics (quadrupolar order) and smectics (layered structure).}
%  we hence dub them ``hybridized nematic order''.

To conclude, we should mention that among all of the energetically degenerate states connected by $\mathbb{Z}_2$ symmetry, three of them have the 16-site cubic unit cell of Fig.~\ref{fig:quadpod_c2dist} paving the entire lattice. They possess two planar symmetries rather than only one, \textit{e.g.} planar $xy$- and $yz$-symmetries with the corresponding weights $w_1=w_3=1, w_2=-1$, \textit{cf.} Table~\ref{table:Bconf}. But any planar spin flip as in Fig.~\ref{fig:planarFlip} would immediately break the fragile cubic symmetry.

%%%%%%%%%%%%%%%%%%%%%%%%%%%%%%%%%%%%%%%%%%%%%%%%%%%%%%%%%%
\section{Relationhips irreps of $T_d$}
%%%%%%%%%%%%%%%%%%%%%%%%%%%%%%%%%%%%%%%%%%%%%%%%%%%%%%%%%%
\label{sec:Td}

TKSVM learns order parameters in the language of the raw data, namely, the ordinary spin degrees of freedom.
To gain more insight, we can cast the machine-learned quantities in terms of the irreducible representations 
of the point group $T_d$~\cite{Yan17a}, which 
\changes{
reflect the human way of thinking about the problem.}

%%%%%%%%%%%%%%%%%%%%%%%%%%%%%%%%%%%%%%%%

\changes{
Following the definitions given in Section~\ref{sec:meet.the.irreps}, 
}
the two order parameters found in the decision function in $d_A^s$ can be expressed 
in terms of irreps of $T_d$ as
\begin{subequations}
\begin{align}
c_1^s &= \f1 4 (\| \vekk{m}_{T_{1-}} \|^2 + \| \vekk{m}_E \|^2) \label{eq:c1mX}\\
c_2^s &= \f1 4 (\| \vekk{m}_{T_{1-}} \|^2 - \| \vekk{m}_E \|^2). \label{eq:c2mX}
\end{align}
\end{subequations}
We see that $c_1^s$ reproduces the ground-state constraint of the rank-2 $U(1)$ gauge theory, but with an additional index $s \in \{ xy, xz, yz \}$ for the spontaneous selection of a spin plane. 
The order parameter $c_2^s$ is nevertheless an emergent quantity that is not evident from direct symmetry arguments. The meaning of the ground-state condition $c_2^s = \pm 1$ now becomes more intuitive in this irrep basis. 
Order by disorder \changes{selects states on $A$--sublattice tetrahedra which transform with 
either $\vekk{m}_{T_{1-}}$ or $\vekk{m}_E$}. 
The alteration of $c_2^s = \pm1$ further means that the system can be viewed as staggered layers of $\vekk{m}_{T_{1-}}$ and $\vekk{m}_E$ $A$-tetrahedra.

%%%%%%%%%%%%%%%%%%%%%%%%%%%%%%%%%%%%%%%%

%In Eq.~\eqref{eq:bondop} below we provide the definition of bond order parameters transforming according to the $A_1$, $E$ and $T_2$ irreps~\cite{Shannon10a}. 
%Here the four vectors $\vekk{S}_{0-3}$ refer to the four spins forming a $B$-tetrahedron.

%%%%%%%%%%%%%%%%%%%%%%%%%%%%%%%%%%%%

\changes{
For the $B$--sublattice tetrahedra, it is more convenient to work with bond--based irreps discussed 
in~\cite{Shannon10a} 
\begingroup
\renewcommand*{\arraystretch}{1.2}
\begin{equation}
\resizebox{0.5\textwidth}{!}{$
\begin{pmatrix}\Lambda_{A_1}\\ \Lambda_{E,1}\\ \Lambda_{E,2}\\ \Lambda_{T_2,1}\\ \Lambda_{T_2,2}\\ \Lambda_{T_2,3}
\end{pmatrix} = 
\begin{pmatrix}
\f{1}{\sqrt{6}} &\f{1}{\sqrt{6}} &\f{1}{\sqrt{6}} &\f{1}{\sqrt{6}} &\f{1}{\sqrt{6}} &\f{1}{\sqrt{6}}\\
\f{1}{\sqrt{3}} &-\f{1}{2\sqrt{3}} &-\f{1}{2\sqrt{3}} &-\f{1}{2\sqrt{3}} &-\f{1}{2\sqrt{3}} &\f{1}{\sqrt{3}}\\
0 &\f{1}{2} &-\f{1}{2} &-\f{1}{2} &\f{1}{2} &0\\
0 &0 &-\f{1}{\sqrt{2}} &\f{1}{\sqrt{2}} &0 &0\\
0 &-\f{1}{\sqrt{2}} &0 &0 &\f{1}{\sqrt{2}} &0\\
-\f{1}{\sqrt{2}} &0 &0 &0 &0 &\f{1}{\sqrt{2}}\\
\end{pmatrix}
\begin{pmatrix}
\vekk{S}_0\cdot\vekk{S}_1\\ \vekk{S}_0\cdot\vekk{S}_2\\ \vekk{S}_0\cdot\vekk{S}_3\\
\vekk{S}_1\cdot\vekk{S}_2\\ \vekk{S}_1\cdot\vekk{S}_3\\ \vekk{S}_2\cdot\vekk{S}_3
\end{pmatrix}
$} \; ,
\label{eq:bondop}
\end{equation}
\endgroup
where the convention for numbering sites is defined in Appendix~\ref{sec:Def}.  
Once transcribed in this basis, the decision function $d_B$ becomes}
\begin{equation}
\label{eq:dBtrans}
d_B\sim \big[a_{A_1} \Lambda_{A_1} + a_{E,1}\Lambda_{E,1} + a_{E,2}\Lambda_{E,2}\big]^2, 
\end{equation}
\changes{with coefficients}
\begin{align}
a_{A_1}&=\sqrt{\frac{2}{3}}\ (w_1+w_2+w_3)\\
a_{E,1}&=\sqrt{\frac{1}{3}}\ (2w_1-w_2-w_3)\\
a_{E,2}&=w_2-w_3.
\end{align}
Maximizing $d_B$, \changes{subject to the constraint of fixed spin length, 
requires $\Lambda_{A_1} \equiv -\sqrt{\f{2}{3}}$ and that $\vekk{\Lambda}_E$ 
take on one of three possible representations}
\begingroup
\renewcommand*{\arraystretch}{1.2}
\begin{equation}
\vekk{\Lambda}^{\star 1}_{E}=\begin{pmatrix}\f{4}{\sqrt{3}}\\ 0\end{pmatrix}\quad
\vekk{\Lambda}^{\star 2}_{E}=\begin{pmatrix}\f{-2}{\sqrt{3}}\\ 2\end{pmatrix}\quad
\vekk{\Lambda}^{\star 3}_{E}=\begin{pmatrix}\f{-2}{\sqrt{3}}\\ -2\end{pmatrix} \; ,
\label{eq:LambdaE}
\end{equation}
\endgroup
which are the three maxima of $\|\vekk{\Lambda}_E\|^2$ under the condition of minimal $\Lambda_{A_1}$.

\begin{figure}[t]
  \centering
  \subfloat[\label{subfig:LambdaE}]{\includegraphics[width=0.45\columnwidth]{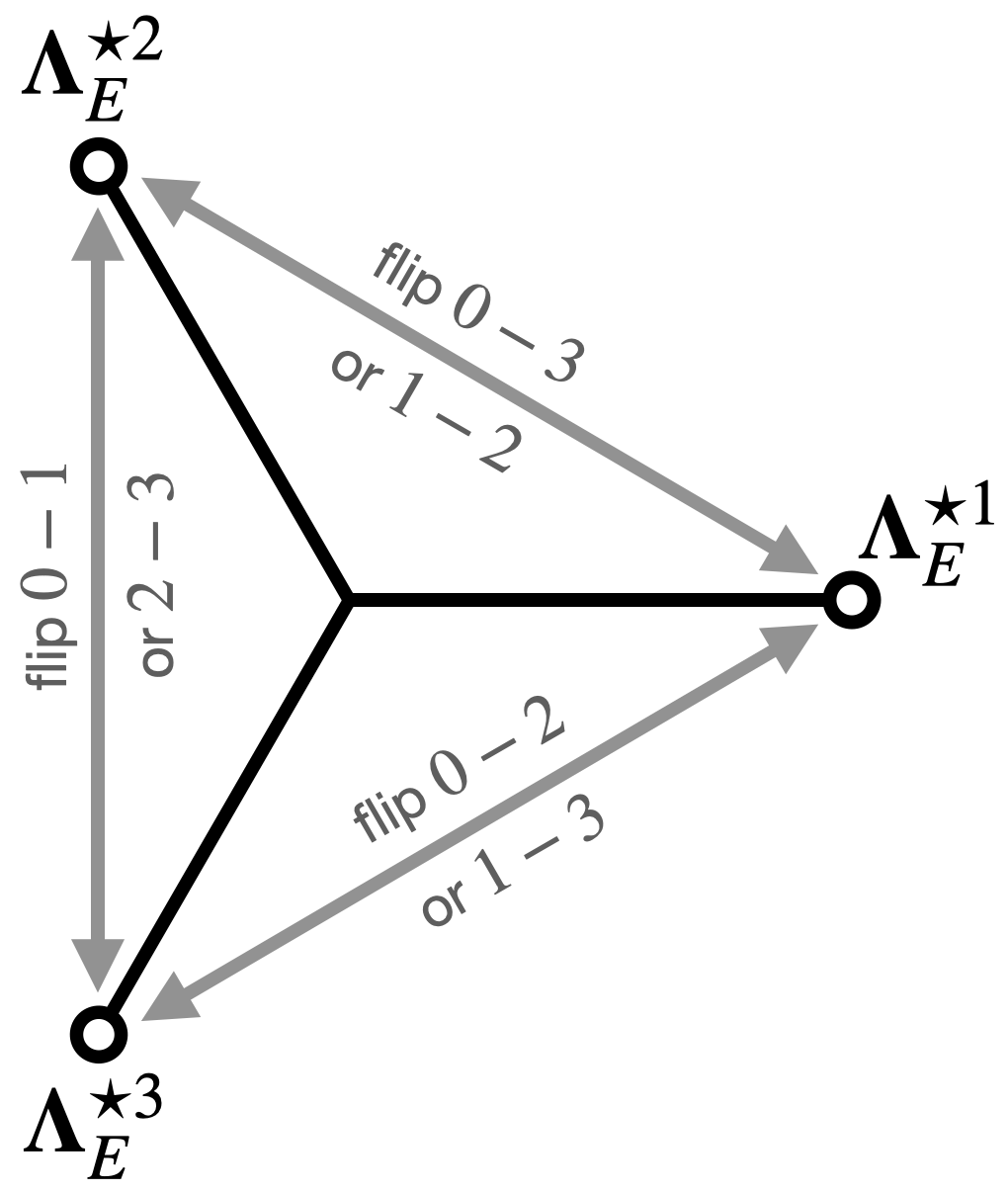}}\\
  \subfloat[\label{subfig:Bflip}]{\includegraphics[width=0.9\columnwidth]{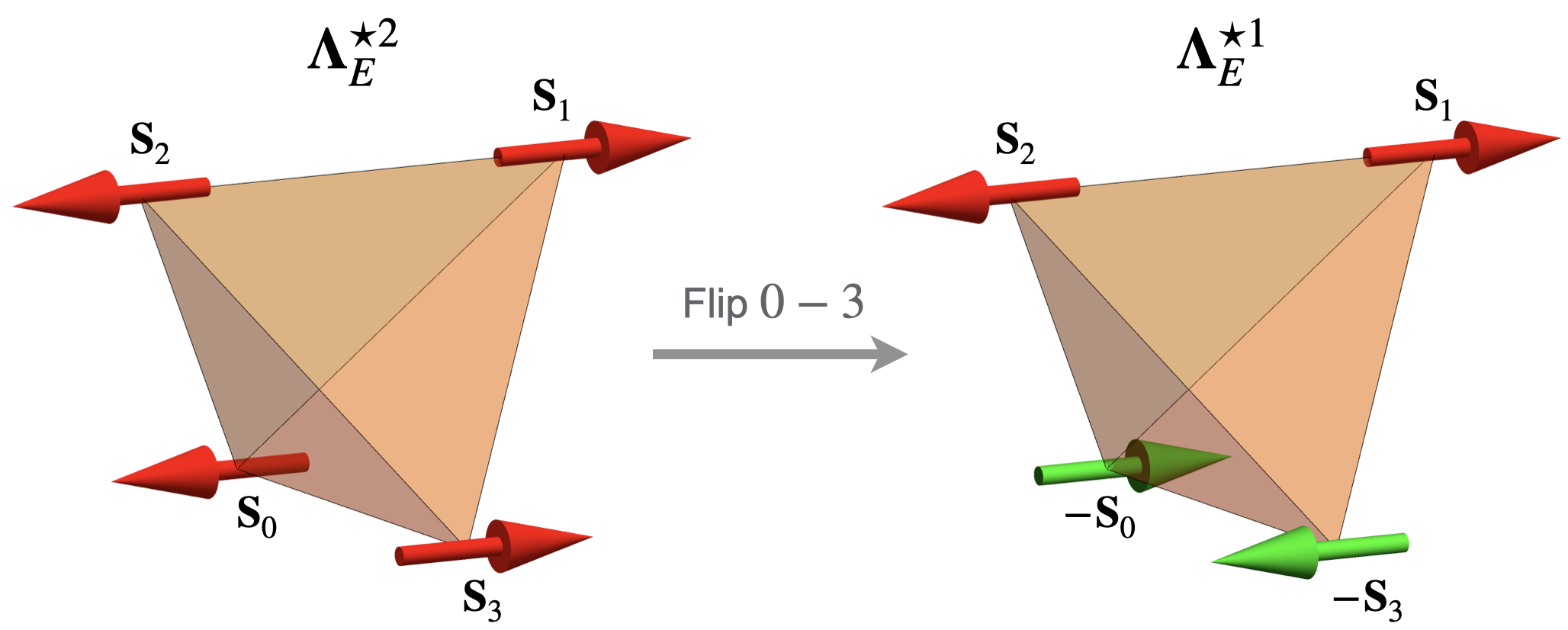}}
  \caption{
  \changes{
  Bond--based irreps of the tetrahedral symmetry group $T_d$ appearing in the decision function 
  found in analysis of the phase with undetermined order.
  (a) Relationship between representations which maximize $\|\vekk{\Lambda}_E\|^2$ [Eq.~(\ref{eq:LambdaE})]
  under $\mathbb{Z}_2$ bond--flip operations [Eq.~(\ref{eq:Z2.symmetry})].
  (b) Illustration of action of $\mathbb{Z}_2$ bond--flip operation on a specific pair of representations, 
  showing how one configuration maximizing $\|\vekk{\Lambda}_E\|^2$ is mapped onto another.
  }
  Bonds with anti-parallel spins are highlighted in green.}
  \label{fig:Bsymm}
\end{figure}

These three configurations are transformed by the $\mathbb{Z}_2$ planar symmetry, as depicted in Fig.~\ref{fig:Bsymm}.
Nevertheless, we can infer the distribution of $\vekk{\Lambda}_E$,
$\{P(\vekk{\Lambda}_E^{\star1}), P(\vekk{\Lambda}_E^{\star2}), 
P(\vekk{\Lambda}_E^{\star3})\}$, 
over all the $B$-tetrahedra from the weights of $d_B$ in Eq.~\eqref{eq:dBorig}.
Specifically, we denote $\vekk{a}^{\star1}_{E}, \vekk{a}^{\star2}_{E}, \vekk{a}^{\star3}_{E}$ to be the respective coefficients in the extreme cases where all $B$-tetrahedra are in the same $\vekk{\Lambda}_E$ configuration, 
\begingroup
\renewcommand*{\arraystretch}{1.2}
\begin{equation}
\vekk{a}^{\star1}_{E}
=\begin{pmatrix}\f{-4}{\sqrt{3}}\\ 0\end{pmatrix},\quad
\vekk{a}^{\star2}_{E}
=\begin{pmatrix}\f{2}{\sqrt{3}}\\ -2\end{pmatrix}\ \ \text{or} \ \
\vekk{a}^{\star3}_{E}
=\begin{pmatrix}\f{2}{\sqrt{3}}\\ 2\end{pmatrix}.
\end{equation}
\endgroup
In addition, $a_{A_1}=\sqrt{2/3}$ reflects the ground state condition $\Lambda_{A_1}=-\sqrt{2/3}$ which is independent of $\vekk{\Lambda}_E$.
The general coefficients $a_{A_1}, a_{E,1}, a_{E,2}$ are then solved from a set of linear equations 
\begingroup
\renewcommand*{\arraystretch}{1.3}
\begin{equation}
\begin{pmatrix}
a_{A_1} & a_{A_1} & a_{A_1} \\
a_{E,1}^{\star1} & a_{E,1}^{\star2} & a_{E,1}^{\star3}\\
a_{E,2}^{\star1} & a_{E,2}^{\star2} & a_{E,2}^{\star3}
\end{pmatrix}
\begin{pmatrix}
P(\vekk{\Lambda}_E^{\star1})\\
P(\vekk{\Lambda}_E^{\star2})\\
P(\vekk{\Lambda}_E^{\star3})
\end{pmatrix}
=
\begin{pmatrix} a_{A_1}\\ a_{E,1}\\ a_{E,2} \end{pmatrix}.
\end{equation}
\endgroup
Here the first equation reduces to the normalisation of the distribution $\sum_i P(\vekk{\Lambda}_E^{\star i})=1$, 
equivalent to $\sum_i w_i = 1$.
In the example of Fig.~\ref{fig:BtetrPattern}, the weights are given by $w_1=-w_2=0.4$ and $w_3=1$ which translates to
$a_{A_1}=\sqrt{2/3}\text{, }a_{E,1}=\sqrt{1/3}\cdot0.2 \text{ and }a_{E,2}=-1.4$. Solving the linear system for these values yields

\begin{equation}
P(\vekk{\Lambda}_E^{\star1})=0.3\quad P(\vekk{\Lambda}_E^{\star2})=0.7\quad P(\vekk{\Lambda}_E^{\star3})=0.
\end{equation}
In general there is always one vanishing $P(\vekk{\Lambda}_{E}^{\star i})$, which is equivalent to have maximal weight $\max w_i = 1$ and can be associated with the spatial orientation of the planar $\mathbb{Z}_2$ symmetry, as listed in Table~\ref{table:Bconf}.

%%%%%%%%%%%%%%%%%%%%%%%%%%%%%%%%

\changes{
From this analysis is clear that the decision function constructed by the SVM in its attempt to classify data
for the unknown ordered phase, has a natural interpretation in terms of physical quantities, namely the 
symmetries of a tetrahedron within the pyrochlore lattice.
In a simpler problem, the irrep corresponding to the ordered state might have been deduced, or even intuited,  
by a human researcher.
However in the present case, many different irreps are involved, and even the form of irrep needed 
changes from A-- to B--sublattice tetrahedra.
At this level of complexity, it would be very hard for a human researcher to arrive at the 
combination of irreps required to classify the unkown order without the guiding hand of the machine.
}

%------------------------------------------------------------------------
%------------------------------------------------------------------------

%%%%%%%%%%%%%%%%%%%%%%%%%%%%%%%%%%%%%%%%%%%%%%%%%%%%%%%%%%
\section{Summary and Conclusions}
%%%%%%%%%%%%%%%%%%%%%%%%%%%%%%%%%%%%%%%%%%%%%%%%%%%%%%%%%%
\label{sec:conclusion}

Although artificial intelligence (AI) brings the opportunity to automate many of the routine, repeated, tasks which 
arise in scientific research, it remains an open question how far AI will impact on the creative, 
conceptual, and problem-solving aspects of science. 
Here, we have shown how researchers trying to understand a difficult problem in frustrated magnetism, 
and faced with poorly--thermalised Monte Carlo simulations, benefit from working with AI.

%%%%%%%%%%%%%%%%%%%%%%%%%%%%%%%%%%%%%%%%%%%%%%%%%%%%%%%%%%

The problem considered was the form of low--temperature order achieved in a model of breathing pyrochlore 
magnets [Eq.~(\ref{eq:H})], as the result of a phase transition out of classical spin liquid 
described by a higher--rank gauge theory (cf. Section~\ref{sec:model}).   
Earlier Monte Carlo (MC) simulations [Yan {\it et al.}, Ref.~\onlinecite{Yan20a}], were sucessful in characterizing the 
spin liquid, and identified a phase transition into an ordered state at very low temperature.
However, simulation results at the lowest temperatures were too poorly-thermalized for 
the form of order to be identified by conventional means (cf. Section~\ref{sec:MC}).

%%%%%%%%%%%%%%%%%%%%%%%%%%%%%%%%%%%%%%%%%%%%%%%%%%%%%%%%%%

In this article, we have demonstrated how a highly--interpretable form of machine learning, 
the support vector machine with tensorial kernel (TKSVM) can be used to re-analyze the same noisy MC data, 
gaining critical insights into the underlying correlations, without the need for prior training or supervision.   
By using kernels based on both rank--1 and rank--2 tensors, we obtain information 
about both dipolar and quadrupolar correlations within spin configurations at low temperatures.

%%%%%%%%%%%%%%%%%%%%%%%%%%%%%%%%%%%%%%%%%%%%%%%%%%%%%%%%%%

Considering first the rank--1 tensor [Section~\ref{sec:rank1}], we found that MC results are inconsistent 
with conventional magnetic order, confirming that a more complex mechanism is at work.
Turning to the rank--2 tensor [Section~\ref{sec:rank2}], we learn that there are three global spin planes, 
$xy$, $xz$, and $yz$ which spontaneously break the spin rotation symmetry on the $A$--sublattice tetrahedra. 
These correlations are captured by a local order parameter, $c_1$ [Eq.~(\ref{eq:definition-c1})], 
%defined on $A-$tetrahedra, 
which selects a (sub-)manifold of the degenerate states contributing to the spin liquid, 
characterized by an emergent $U(1)$ degree of freedom [Eq.~(\ref{eq:Aconfxy})].  
A second local order parameter $c_2$ [Eq.~(\ref{eq:definition-c2})] further selects four discrete states from this $U(1)$ manifold 
%captures the effects of fluctuations involving $B$--subblattice tetrahedra
[Eq.~(\ref{eq:spThetas})].

Applying TKSVM to $B$--sublattice tetrahedra then brings forward the full three-dimensional magnetic structure, i.e. how these four discrete states pave the entire lattice together.

%%%%%%%%%%%%%%%%%%%%%%%%%%%%%%%%%%%%%%%%%%%%%%%%%%%%%%%%%%

We then returned to MC simulation, armed with what had been 
learned through the application of machine learning \mbox{[Section~\ref{sec:MC2}]}.  
By generating spin configurations that satisfy the constraints imposed by the local 
order parameters $c_1$ and $c_2$, we are able to initialize MC simulations 
within the manifold of states favoured at low temperatures. 
Doing so, we found that simulations converge nicely [Fig.~\ref{fig:MC2}], 
confirming the validity of the machine--learning analysis, and leading to a 
consistent estimate of the ordering temperature.

%%%%%%%%%%%%%%%%%%%%%%%%%%%%%%%%%%%%%%%%%%%%%%%%%%%%%%%%%%

As well as facilitating improved simulations, machine learning results also teach us about the deeper 
structure of the manifold of states favoured at low temperatures.
In Section~\ref{sec:Z2symmetry} we revisited the decision function found by the TKSVM, 
and show how it is invariant under operations which ``flip'' all spins within an entire layer of A--sublattice 
tetrahedra.
From this we infer the existence of an emergent $\mathbb{Z}_2$ symmetry, associated with  
planes of tetrahedra.
This leads to a picture which has much in common with a smectic liquid crystal, 
in which translational symmetries are (partially) broken by parallel planes of molecules
that align to common axis.
%
% In the context of the breathing pyrochlore lattice, we dub this a ``\lode{hybridized} nematic order''.

%%%%%%%%%%%%%%%%%%%%%%%%%%%%%%%%%%%%%%%%%%%%%%%%%%%%%%%%%%

Finally, we considered how machine learning results might be interpreted in terms of 
a more conventional language of symmetry operations [Section~\ref{sec:Td}].  
We found that the local order parameter $c_1$ could be expressed in terms of the irreps of the symmetry group 
of a tetrahedron, $T_d$ [Appendix~\ref{sec:Def}].
Comparing this with the theory of the spin liquid found at higher temperature \cite{Yan20a}, we learned that 
the low temperature order satisfies the emergent Gauss' law which defines the spin liquid, 
and can be associated with fluctuation of the electric field tensor of the associated higher--rank gauge theory.

%There is in addition a remaining planar $\mathbb{Z}_2$ subsystem symmetry where flipping all spins on a $A$-layer leaves its $c_2$ value invariant. The ground state is thus a rare instance of what we call a ``hybridized nematic order'', with the co-existence of dipolar (rank-1) and quadrupolar (rank-2) orders due to an emergent subdimensional symmetry.\\

%%%%%%%%%%%%%%%%%%%%%%%%%%%%%%%%%%%%%%%%%%%%%%%%%%%%%%%%%%

These results are remarkable in a number of ways.
Firstly, they demonstrate the power of the machine learning approach in extracting meaningful 
physical information from noisy MC data. 
Despite the fact that low--temperature simulations were of too poor quality to be interpreted 
by human researchers, the decision function extracted from TKSVM analysis contains sufficient 
information to reconstruct fluctuations of a rank--2 electric field.
This is even more surprising, given that the result was obtained without any prior training or supversion:
at no point was the AI ``taught'' about existence of a tensor electric field within the parent spin liquid phase, 
let alone its possible role in low--temperature order.

%%%%%%%%%%%%%%%%%%%%%%%%%%%%%%%%%%%%%%%%%%%%%%%%%%%%%%%%%%

In rediscovering a key organizational principle, and pinpointing how this relates to an unidentified 
form of low--temperature order, the AI is performing a role which would usually be associated with 
the creative thinking of a human researcher.
But this is not to say that AI entirely replaced, or superceeded, human insight.
% in the solution of this problem.
%
Instead, the analysis of the decision function found through the TKSVM provided the ``missing link'' 
needed for a team of human researchers to resolve the remaining open questions.

%%%%%%%%%%%%%%%%%%%%%%%%%%%%%%%%%%%%%%%%%%%%%%%%%%%%%%%%%%

In this respect, the input from AI played exactly the same role as the input from 
any human researcher who provides a critical insight into an unsolved problem.   
Ultimately, the solution was accomplished by combining the input 
from AI with human insights into breathing--pyrochlore model, and through 
new MC simulations seeded from spin configurations generated using the help 
of the TKSVM. 
This interplay of ideas generated by humans, and ideas (or more precisely, equations) 
generated by a machine, follows the familiar pattern of collaboration within a team 
of researchers, each of whom makes their own, interdependent, contribution to the 
solution of a problem.

%%%%%%%%%%%%%%%%%%%%%%%%%%%%%%%%%%%%%%%%%%%%%%%%%%%%%%%%%%

Within the field of frustrated magnetism, there any many other problems which might benifit from such an approach.
In our case, it was crucial to have a very strongly interpretable algorithm, but  analytical and 
group-theoretical arguments were also indispensable. 
We do not expect all noisy simulations to be tractable this way. For instance, the critical slowing down 
witnessed in second-order phase transitions would be especially challenging, since it requires updates acting 
at different length scales. 
That being said, our approach is sufficiently general that it can be applied to a variety of complex unknown phases. 
What immediately comes to mind are other frustrated magnets such as the Gd 
pyrochlores~\cite{javanparast15a,Welch22a} and Mn-kagome \cite{Paddison23a} with multi$-k$ orders, 
the subtle ordering of dipolar pyrochlores \cite{palmer00a,McClarty14a} or Kitaev 
magnets \cite{Rau14a,janssen19a,rao21a}. 
As the TKSVM algorithm has also been extended to quantum problems~\cite{Sadoune23}, our strategy 
is not confined to classical physics, and could even be combined with the field of quantum simulation. 

%%%%%%%%%%%%%%%%%%%%%%%%%%%%%%%%%%%%%%%%%%%%%%%%%%%%%%%%%%

More generally, the ``collaborative'' nature of the interactions between human researchers and machines 
described in this paper become possible wherever the output of machine--learning can be translated into a form 
which can be understood and manipulated by humans, such as the equations implied by the decision function 
of a support vector machine.
While the TKSVM is ideally suited to such an approach, it is not the only interpretable form of machine learning, 
and even approaches that are not directly interpretable, such as deep neural networks, may provide 
critical insights into unsolved problems.   
As such this model of collaboration offers one possible paradigm for AI-driven research in an age in which 
human and machine have complementary strengths.

%%%%%%%%%%%%%%%%%%%%%%%%%%%%%%%%%%%%%%%%%%%%%%%%%%%%%%%%%%
\begin{acknowledgements}
N.S., K.L., and L.P. acknowledge support from FP7/ERC Consolidator Grant QSIMCORR, No. 771891, and the Deutsche Forschungsgemeinschaft (DFG, German Research Foundation) under Germany's Excellence Strategy -- EXC-2111 - 390814868.
L.D.C.J. acknowledges financial support from CNRS (PICS France-Japan MEFLS) and from the French "Agence Nationale de la Recherche" under Grant No. ANR-18-CE30-0011-01. 
K.L. acknowledges support from the New Cornerstone Science Foundation through the XPLORER PRIZE, Anhui Initiative in Quantum Information Technologies, and Shanghai Municipal Science and Technology Major Project (Grant No. 2019SHZDZX01).
N.S. acknowledges financial support from the Theory of Quantum Matter Unit, OIST, 
and JSPS KAKENHI Grants No. JP19H05822 and JP19H05825.  
%
%L.P. acknowledge financial support from ANR-DFG.
\end{acknowledgements}

%%%%%%%%%%%%%%%%%%%%%%%%%%%%%%%%%%%%%%%%%%%%%%%%%%%%%%%%%%
\begin{appendix}
%%%%%%%%%%%%%%%%%%%%%%%%%%%%%%%%%%%%%%%%%%%%%%%%%%%%%%%%%%
	
\section{Definitions and Conventions}
\label{sec:Def}

%In this Appendix we provide the definitions of the breathing pyrochlore lattice, the shape of the DM interactions, and explicit expressions for the order parameters breaking the point-group symmetry of a single  tetrahedron with fields transforming as the  irreps $\{ A_2, E, T_2, T_{1+}, T_{1-} \} $ of the $T_d$ group. \\
In this Appendix we provide lattice and model definitions, and our conventions for the order parameters of the irreps of the $T_d$ group.

The sites of an $A$-tetrahedron are located at

\begin{alignat*}{2}
&\vekk{r}_0=\frac{a}{8}(1,1,1)\quad
&&\vekk{r}_1=\frac{a}{8}(1,-1,-1)\\
&\vekk{r}_2=\frac{a}{8}(-1,1,-1)\quad
&&\vekk{r}_3=\frac{a}{8}(-1,-1,1)
\end{alignat*}

relative to the center of an $A$-tetrahedron, as in~\cite{Ross11a}. 
Here, $a$ is the length of the FCC unit cell.
The sites of a $B$-tetrahedron are located at $-\vekk{r}_0,-\vekk{r}_1,-\vekk{r}_2,-\vekk{r}_3$
relative to the center of a $B$-tetrahedron.\\

The bond dependent DM-interaction vectors are defined as \cite{Kotov05a,canals08a}

\begin{alignat*}{3}
&\vekk{d}_{01}=\frac{(0,-1,1)}{\sqrt{2}}\
&&\vekk{d}_{02}=\frac{(1,0,-1)}{\sqrt{2}} \
&&\vekk{d}_{03}=\frac{(-1,1,0)}{\sqrt{2}} \\
&\vekk{d}_{12}=\frac{(-1,-1,0)}{\sqrt{2}} \
&&\vekk{d}_{13}=\frac{(1,0,1)}{\sqrt{2}} \
&&\vekk{d}_{23}=\frac{(0,-1,-1)}{\sqrt{2}}
\end{alignat*}

We do not consider models with DM couplings $D_B$ on B tetrahedra because $D_B$ lifts the degeneracy of the rank$-2$ spin liquid, thus stepping away from our motivation to study the ordering mechanism within a higher-rank gauge theory. Nevertheless, even if it can be expected to be much smaller than on $A-$tetrahedra $D_B \ll D_A$, $D_B$ is allowed by symmetry. If $D_B<0$, then the system should ultimately order the same way as for non-breathing pyrochlores, i.e. via order by disorder within $\Gamma_5$ states \cite{canals08a,chern10b,noculak23a}. If $D_B>0$, there would be an energetic competition between $\Gamma_5$ and all-in/all-out states depending on the ratio $D_A/D_B$. Note that if $0< |D_B| \ll T_c$ then we should recover the physics discussed in this paper at $T_c$ before a second phase transition at lower temperatures.

%In Table~\ref{table:irreps.Td} we provided explicit expressions for the order parameters breaking the point-group symmetry of a single  tetrahedron~\cite{Yan20a,Yan17a}, used to describe learned TKSVM order parameters on an $A$-tetrahedron.

%%%%%%%%%%%%%%%%%%%%%%%%%%%%%%%%%%%%%%%%
\section{TKSVM}
%%%%%%%%%%%%%%%%%%%%%%%%%%%%%%%%%%%%%%%%
\label{sec:TKSVM.overview}

This section aims at providing the reader with the essential working principles of the tensorial-kernel support vector machine (TKSVM). The TKSVM approach is an interpretable and (quasi-)unsupervised machine learning algorithm developed in Refs.~[\onlinecite{Liu19a},\onlinecite{Greitemann19b},\onlinecite{Greitemann19a}] and recently also extended to quantum problems~\cite{Sadoune23}.

Considering classical $O(3)$-spin configurations (ie, Monte Carlo snapshots) $\mathbf{x} = \{S_i^a | i = 1, 2, ..., N; a = x,y,z \}$, the first step of TKSVM is the construction of feature vectors $\bds{\phi} = \{\phi_\mu\}$ consisting of degree-$n$ monomials from $\mathbf{x}$
\begin{align}\label{eq:phi}
    \phi_{\mu} = \corr{S_{\alpha_1}^{a_1} S_{\alpha_2}^{a_2} \dots S_{\alpha_n}^{a_n}}_{\rm cl},
\end{align}
where $\corr{\cdots}_{\rm cl}$ represents a lattice average over pre-determined non-overlapping clusters, each containing $r$ spins, where $\alpha_1, \dots, \alpha_n$ label spins within the cluster, and where $\mu = \{\alpha_1, a_1; \dots; \alpha_n, a_n \}$ denotes a composite index. The tensorial feature space spanned by $\{\phi_{\mu}\}$ hosts any potential classical spin-order of degree $n$ that fits within the pre-defined cluster of size $r$. Following the construction of feature vectors from the input data, TKSVM detects the underlying order during the learning stage, provided that the user made a suitable choice of the hyper-parameters $n$ and $r$. The optimal choice of $n$ and $r$ are unknown a priori. Therefore we choose clusters in accordance with the unit-cell of the lattice and Hamiltonian interactions, and increase $n$ systematically on a trial-and-error basis until TKSVM succeeds. In this approach $n = 1$ allows the detection of magnetic orders, and higher $n > 1$ detects multipolar orders and emergent local constraints. 

A central concept of SVM methods is the decision function $d$.  The decision function can be written as a product $d=\mathbf{V}^t \mathbf{\hat{C}}\mathbf{V}$ up to a constant known as the bias. Here, $\mathbf{V}$ is a vector made of input data and $\mathbf{\hat{C}} = \{C_{\mu\nu}\}$ is the output of the machine in the form of a coefficient matrix  measuring correlations of $\phi_{\mu}$,
\begin{align}\label{eq:C_munu}
    C_{\mu \nu} = \sum_k \lambda_k \phi_\mu(\mathbf{x}^{(k)}) \phi_\nu(\mathbf{x}^{(k)}),
\end{align}
where the Lagrange multiplier $\lambda_k$ denotes the weight of the $k$-th sample.
The non-vanishing entries of $C_{\mu\nu}$ identify the relevant basis tensors of the tensorial feature space, and their interpretation yields analytical expressions of the underlying order parameters.
As TKSVM has never learned nor seen any of the different phases, it has the advantage of being unbiased in identifying them. Compared to a human approach where one would define an order parameter and then test it on the Monte Carlo data, the approach in TKSVM is blind to specifying order parameter candidates and looks for all possibilities within the search space spanned by the rank-$r$ monomials defined on the cluster of size $n$.

%%%%%%%%%%%%%%%%%%%%%%%%%%%%%%%%%%%%%%%%%
\section{Pattern Interpretation}
%%%%%%%%%%%%%%%%%%%%%%%%%%%%%%%%%%%%%%%%%
\label{sec:PatternInterpretation}

The last step of TKSVM consists of constructing the analytical expression of the underlying order from the internal parameters of the learning model. This is achieved by reading off and interpreting the graphical representation (pattern) of the coefficient matrix. Since the underlying order in this case is fairly complex, we shall discuss the procedure for a subset of the full pattern for the $A$-tetrahedra only. Specifically, we  start by considering the block with spin indices $(2\,3,3\,0)$; see the upper zoomed-in panel of Fig.~\ref{fig:AtetrPattern}.

Reading off the terms from the block pattern with coefficients approximated as $\pm 1$ yields the expression
\begin{equation}
\begin{split}
&[d_A]_{(2\,3,3\,0)}\sim \\
&+(S_2^y S_3^x)(S_3^x S_0^x)
+(S_2^y S_3^x)(S_3^y S_0^y)
-(S_2^x S_3^y)(S_3^x S_0^x)\\
&-(S_2^x S_3^y)(S_3^y S_0^y)
+(S_2^y S_3^y)(S_3^x S_0^y)
+(S_2^y S_3^y)(S_3^y S_0^x)\\
&-(S_2^x S_3^x)(S_3^x S_0^y)
-(S_2^x S_3^x)(S_3^y S_0^x).
\end{split}
\end{equation}
In order to reshape the expression into a sum over square magnitudes of rank-2 order parameters, we factorize the feature components and assign the coefficients (signs) in consistency with other block patterns $(2\,3,2\,3)$, $(3\,0,3\,0)$ and $(3\,0,2\,3)$
\begin{equation}
\begin{split}
&[d_A]_{(2\,3,3\,0)} + [d_A]_{(3\,0,2\,3)} + [d_A]_{(2\,3,2\,3)} + [d_A]_{(3\,0,3\,0)}\sim \\
&(-S_2^y S_3^x  +S_2^x S_3^y  -S_3^x S_0^x  -S_3^y S_0^y)^2\\
&+(+S_2^x S_3^x  -S_2^y S_3^y  -S_3^x S_0^y  -S_3^y S_0^x)^2.
\end{split}
\end{equation}
We factorize even further, which requires to consider some more block patterns 
\begin{equation}
\begin{split}
&\sum_{ij,i'j'\in\{0,2,3\}} [d_A]_{(i\,j,i'\,j')} \sim \\
&-\big((S_0^x -S_2^y -S_3^x)^2 + (S_0^y +S_2^x -S_3^y)^2\big)^2 \\
&-\big(2(S_0^x -S_2^y -S_3^x)\cdot(S_0^y +S_2^x -S_3^y)\big)^2.
\end{split}
\end{equation}
This expression already contains a substantial part of the full decision function. Comparing to the definition of $c_1^{xy}$ and $c_2^{xy}$ in Eqs.~\eqref{eq:definition-c1} and~\eqref{eq:definition-c2}, respectively, reveals that only the terms including $\mathbf{S}_1$ are missing.\\
Extending the interpretation to the full pattern, we arrive at the expression
\begin{equation}
d_A = \sum_{\substack{ij,i'j'\\ \in\{0,1,2,3\}}} [d_A]_{(i\,j,i'\,j')} \sim 
-\big((c_1^{xy})^2 + (c_2^{xy})^2\big).
\end{equation}
Note that the overall minus sign is of technical origin and is arbitrary in each TKSVM run, hence it can be dropped. The sign convention in Eq.~\eqref{eq:dA} is chosen to match with the signs in the existing definition of $\mathbf{m}_{T_{1-}}$, see its third component in Table~\ref{table:irreps.Td} for comparison. Furthermore a factor of $1/16$ was introduced to normalize the order parameters to $1$ when saturated (deep in phase).

\end{appendix}

\bibliography{paper.bib}

\end{document}